\documentclass[twocolumn,trackchanges]{aastex631}

\usepackage{mathptmx}
\usepackage{mathrsfs}
\usepackage{xcolor}
\usepackage[mathscr]{euscript}
\usepackage{hyperref}
\usepackage{physics}
\usepackage{simplewick}
\usepackage[bottom]{footmisc}
\usepackage{appendix}
\usepackage{amssymb}
\usepackage{amsmath}
\usepackage{booktabs} % 导入三线表需要的宏包
\usepackage{threeparttable}
\usepackage{threeparttablex}
\usepackage{rotating}
\usepackage{longtable}
\usepackage{multirow}

\newcommand{\R}[1]{\textcolor{red}{#1}}
\newcommand{\B}[1]{\textcolor{blue}{#1}}
\newcommand{\be}{\begin{equation}}
\newcommand{\ee}{\end{equation}}

\newcommand{\bs}{\begin{split}}
\newcommand{\es}{\end{split}}
\graphicspath{{./}{figure/}}
\usepackage[T1]{fontenc}
\usepackage{subfigure}
\usepackage{longtable,booktabs,multirow,threeparttablex}
\usepackage{natbib} % 适配\citet命令

\begin{document}
\title{Investigating the Circumnuclear Medium of Tidal Disruption Events with Radio Observations}
%\title{Comprehensive Study of the Circumnuclear Medium of Tidal Disruption Events with Radio Observations}

\correspondingauthor{Wei-Hua Lei, Shao-Yu Fu}
\email{leiwh@hust.edu.cn, syfu@hust.edu.cn}

\author[0009-0005-9790-1263]{Chang Zhou}
\affiliation{Department of Astronomy, School of Physics, Huazhong University of Science and Technology, Wuhan, 430074, China}

\author[0000-0003-3440-1526]{Wei-Hua Lei}
\affiliation{Department of Astronomy, School of Physics, Huazhong University of Science and Technology, Wuhan, 430074, China}
\affiliation{Hubei Key Laboratory of Gravitation and Quantum Physics, Huazhong University of Science and Technology, Wuhan, Hubei, China}
\affiliation{Hubei Fundamental Research Center for Physics, Wuhan, Hubei, China}

\author[0009-0003-0516-5074]{Xiangli Lei}
\affiliation{Department of Astronomy, School of Physics, Huazhong University of Science and Technology, Wuhan, 430074, China}

\author[0009-0000-0635-5679]{Po Ma}
\affiliation{Department of Astronomy, School of Physics, Huazhong University of Science and Technology, Wuhan, 430074, China}

\author[0009-0002-7730-3985]{Shao-Yu Fu}
\affiliation{Department of Astronomy, School of Physics, Huazhong University of Science and Technology, Wuhan, 430074, China}
\affiliation{Key Laboratory of Space Astronomy and Technology, National Astronomical Observatories, Chinese Academy of Sciences, Beijing, 100101, China}

\author[0000-0002-9022-1928]{Zi-Pei Zhu}
\affiliation{Department of Astronomy, School of Physics, Huazhong University of Science and Technology, Wuhan, 430074, China}
\affiliation{Key Laboratory of Space Astronomy and Technology, National Astronomical Observatories, Chinese Academy of Sciences, Beijing, 100101, China}
 
\begin{abstract}
Tidal disruption events (TDEs) are unique tools for investigating quiescent supermassive black hole (SMBH), accretion physics, and circumnuclear medium (CNM) environments. The CNM density profile is of great astrophysical significance, since it provides key diagnostics for the accretion history of dormant SMBH. TDEs can launch outflows that produce radio emission when propagating into the CNM. The synchrotron radio flux density can be described by a series of power-law segments $F_\nu \propto t^{-\alpha} \nu^{-\beta}$. The closure relation (CR), i.e., the relation between the temporal indices $\alpha$ and the spectral indices $\beta$, are therefore monitoring the CNM density profile. In this work, we first collect 53 TDEs with radio observations to date. We then obtain the predicted CR for arbitrary CNM and different dynamical phases of the outflow, and apply to the radio TDE sample. We constrain the CNM density profile for 26 radio TDEs with good data quality. The results are generally consistent with those estimated with equipatition method, suggesting that CR analysis is efficient in the study of CNM profile for a quiescent SMBH.

%corresponding to different dynamical phases of the outflow, and apply it to 26 out of the 53 collected radio TDEs to constrain their CNM density profiles. %Finally, we compare our results with previous estimations based on equipartition analysis. 

\end{abstract}

\keywords{\href{http://astrothesaurus.org/uat/1696}{Tidal disruption (1696)}}

\section{Introduction} \label{sec:intro}
%tde, environment importance, accretion history
Tidal disruption events (TDEs) occurs when a star passes too close to a SMBH and is torn apart by its intense tidal forces, producing luminous electromagnetic emission \citep{Rees1988,Phinney1989}. It has been observed in various wavelengths (radio, UV/optical, X-rays, even $\gamma$-ray in some jetted TDEs), providing useful tools to investigate quiescent SMBH, accretion physics and CNM environments. 

The CNM density profile ($n\propto R^{-k}$) in quiescent galaxies reflects the accretion history of the central SMBH. Depending on the value of the density profile $k$, the corresponding physical scenarios are as follows: (1) $k=1.5$, consistent with Bondi accretion \citep{bondi1952}; (2) $k\sim 2.5$, indicative of a past episode of super-Eddington accretion \citep{Coughlin2014}; and (3) $k\sim1$, which may suggest replenishment by external stellar winds \citep{Baganoff03}.

Theoretical and observational studies unveiled outflows (e.g., jet, wind, unbound tidal debris) in TDEs \citep{Dai2018,Alexander2020,Dai2021,Qiao2025-simulation}. Radio emissions are produced when the outflow interacts with the CNM. As the outflow expands, the radio emission observed at different epochs remains an independent measurement of the CNM density profile at sub-parsec or even parsec scales, which are not directly resolvable at any wavelength with current facilities at the distance of most TDE hosts \citep{Alexander2020}.

\citet{Alexander2020} reviewed the radio observations of TDEs, focusing on nine TDEs with published radio detections. However, the sample of TDEs with radio detections has increased greatly recently \citep{Cendes2024,Goodwin2025-22eROSAT_sample, Anumarlapudi2024-radiosample}. Some TDEs were previously characterized as non-detected, but have radio detections later on \citep{Alexander2025-delayed_radio_sample}. In this paper, we collect a sample of 53 TDEs with radio detections to date. A comprehensive study on the CNM environments of these radio TDEs are highly desired.

Equipartition analysis has been widely used to estimate the CNM profile of TDEs, which characterizes the system at minimum total energy \citep{Barniol2013}. If the electron and magnetic field energies are close to equipartition then this lower limit is also a good estimate of their true energy. However, the outflow might contain additional components in the total energy (like that carried by protons), and the system would be far from the equipartition. Therefore, an independent method other than equipartition analysis, and a systematical inverstigation of the CNM profile for the full sample of radio TDEs are expected.
 
The synchrotron radio flux density can be described by a series of power-law segments $F_\nu \propto t^{-\alpha} \nu^{-\beta}$ \citep{Sari1998,Gao+2013,Zhang2018}. In such a model, the type of CNM can be tested with the closure relations (CRs, relations between the temporal indices $\alpha$ and spectral indices $\beta$), as did in gamma-ray bursts (GRBs) \citep{Gao+2013}. \citet{Gao+2013} only presented the CRs for constant-density ($k=0$) and wind type ($k=2$). In this work, to study the CNM type of radio TDEs, we obtain the results closure‑relations for arbitrary CNM density profiles as described in several previous works \citep{Eerten2009, Fraija2021}.

The structure of this paper is organized as follows. In Section\,\ref{sec:radio_tde}, we collect the currently known sample of TDEs exhibiting radio emission. In Section\,\ref{sec:model}, we provide the CR formulations corresponding to the interaction of an outflow with an arbitrary CNM density distribution in each dynamical phase, and provide the procedures for determining the spectral index and the temporal decay index. In Section\,\ref{sec:results+discuss}, we report the results of CNM density profile index for a subset of the radio TDE sources with good data quality and discuss the implications of these results. The main conclusions are summarized in Section\,\ref{sec:summary}. A standard cosmology model is adopted with $H_{0}=67.3\ \rm{km\cdot s^{-1} Mpc^{-1}}$, $\Omega_{M}$=0.315, $\Omega_{\Lambda}$=0.685 \citep{Planck+2014}.

\section{Radio-Detected TDE Sample}
\label{sec:radio_tde} 
\begin{figure*}[htp]
\center
\includegraphics[width=1.0\textwidth]{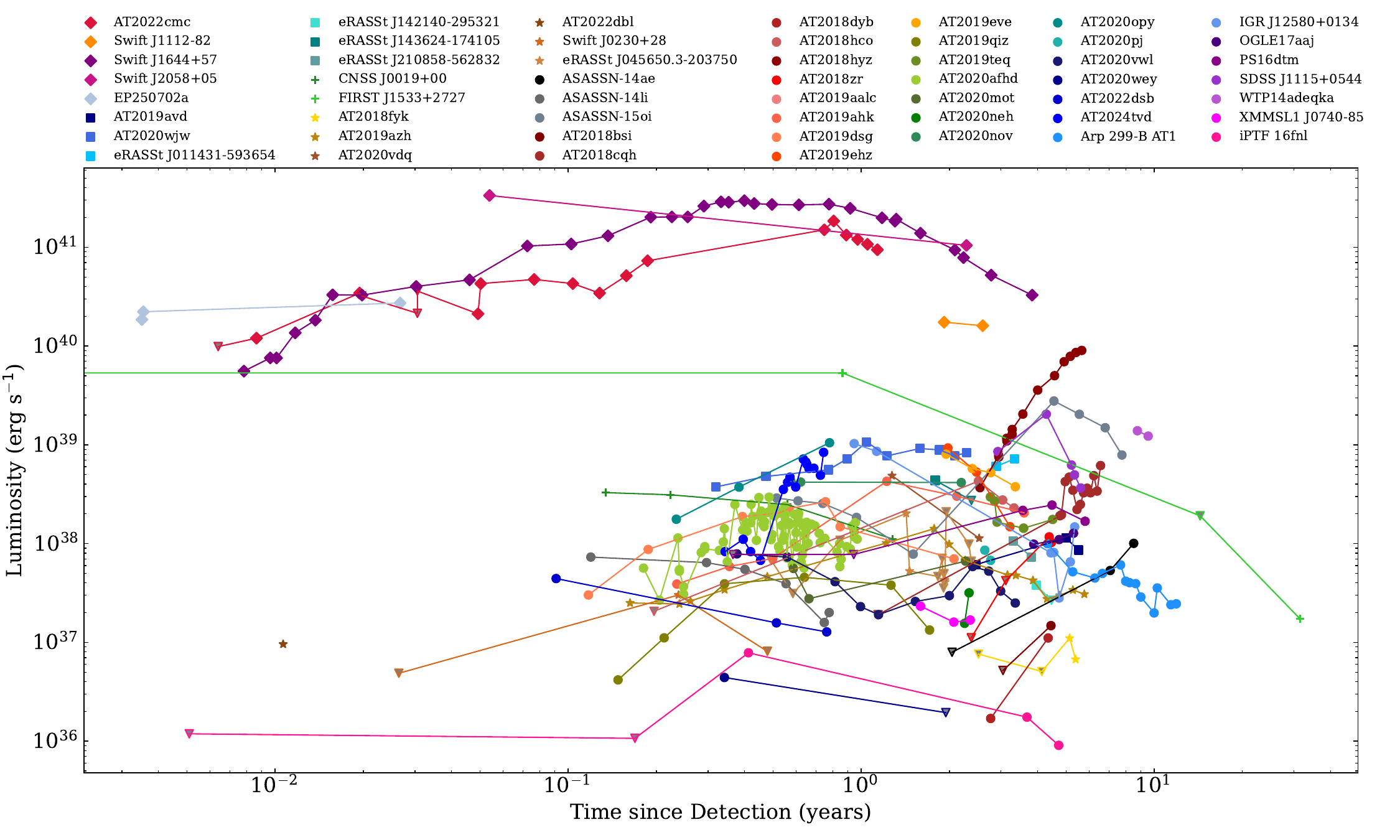}
\caption{The radio light curves for our collected TDEs. To date, 53 TDEs have published radio detections. Although most of the detected TDEs were observed at multiple frequencies, for simplicity we show only a single frequency for each event. The triangles indicate the observations with published upper limits. The data can be available at \url{https://github.com/leiwh/RadioTDEs}.}
\label{fig:radio TDEs}
\end{figure*}

In this section, we present a compilation of TDEs with radio emission known up to early 2026. We first collected the radio data for these sources from previous reports and the literature. In addition, we processed the VLA observations of two sources, AT2019teq and AT2020vdq. This effort resulted in a sample of 53 TDEs with detected radio emission. The information and radio data references for these sources are summarized in Table \ref{tab:all_radio_tde}, and the radio light curves (around 5\,GHz) for these TDEs are plotted in Figure \ref{fig:radio TDEs}.

\subsection{VLA Data Reduction}
In this work, we incorporate additional archival VLA observations of AT2019teq and AT2020vdq obtained from the NRAO Science Data Archive. For AT2019teq, we use observations from programs 24A-322 (PI: Cendes), 20B-377 (PI: Alexander), and 23A-241 (PI: Cendes), together with data products from VLASS 4.1. For AT2020vdq, we analyze VLA observations from programs 23A-409 (PI: Somalwar) and 21B-322 (PI: Hallinan), as well as data products from VLASS 3.2.
The data reduction is carried out using the Common Astronomy Software Applications package (CASA 6.5.4), following standard VLA calibration procedures. The visibility data are calibrated using the automated VLA calibration pipeline provided in CASA. Primary flux density and bandpass calibration were performed using standard calibrators (3C147 or 3C286), depending on the observing epoch, while nearby phase calibrators were used for gain calibration. After pipeline processing, the data were manually inspected and additional flagging was applied to remove radio frequency interference (RFI) and poorly calibrated scans when necessary.
Imaging was performed using the CASA task \texttt{TCLEAN} in multi-scale multi-frequency synthesis (MS-MFS) mode to properly account for the wide fractional bandwidth and possible spectral variations. We adopted Briggs weighting with a robust parameter of 0 or 2 to achieve an optimal compromise between angular resolution and sensitivity. The flux density of AT2019teq and AT2020vdq was measured by fitting a point-source Gaussian model in the image plane using the CASA task \texttt{IMFIT}. The local image rms noise was estimated with \texttt{IMSTAT} in nearby source-free regions. The flux density uncertainty was calculated as the quadratic sum of the \texttt{IMFIT} fitting uncertainty and a typical systematic calibration uncertainty of 5\%. All measured flux densities and upper limits are summarized in Table \ref{tab:AT2019teq+at2020vdq}.

\begin{deluxetable*}{lcccccc}
\tablecaption{Radio observations of AT2019teq and AT2020vdq.\label{tab:AT2019teq+at2020vdq}}
\tablehead{
\colhead{Object} &
\colhead{Telescope} &
\colhead{Project Code / Source} &
\colhead{Date of Observation} &
\colhead{$\delta t$ (d)} &
\colhead{Frequency (GHz)} &
\colhead{Flux Density (mJy)}
}
\startdata
AT2019teq & VLA & 20B-377 & 2022 Nov 26 & 1133  & 5.0  & $0.26\pm0.021$ \\
---      & --- & --- & ---& --- & 6.0  & $0.217\pm0.015$ \\
---      & --- &  ---  &  ---  &  ---  & 7.0 & $0.171\pm0.013$ \\
---      & --- & ---  & ---  & ---  & 9.0 & $0.16\pm0.014$  \\
---      & --- & ---& --- & --- & 11.0  & $0.118\pm0.013$ \\
---      & --- & 23A-241 & 2023 Sep 12 & 1423 & 1.5  & $0.311\pm0.029$ \\
---      & --- &  --- &  --- &  --- & 3.0  & $0.203\pm0.021$ \\
---      & --- & --- & --- & --- & 5.0  & $0.138\pm0.019$ \\
---      & --- & --- & --- & --- & 6.0  & $0.115\pm0.011$ \\
---      & --- & --- & --- & --- & 7.0 & $0.097\pm0.019$ \\
---      & --- & ---& --- & ---& 9.0  & $<0.08$  \\
---      & --- & --- & ---  & ---  & 11.0  & $<0.08$  \\
---      & --- & 24A-322 & 2024 Sep 11 & 1788 & 6.0  & $0.141\pm0.01$ \\
---      & --- & VLASS 4.1 & 2025 Sep 23 & 2165 & 3.0  & $<0.371$ \\
AT2020vdq & --- & VLASS 3.2 & 2024 Jul 9 & 1374 & 3.0 & $<0.381$\\
--- & --- & 21B-322 & 2022 Feb 2 & 486 & 1.5 & $0.953\pm0.09$ \\
--- & --- & --- & --- & ---  & 3.0 & $1.485\pm0.09$\\
--- & --- & --- & --- & ---  & 3.5 & $1.717\pm0.11$\\
--- & --- & --- & --- & ---  & 4.5 & $1.906\pm0.11$\\
--- & --- & --- & --- & ---  & 5.0 & $1.897\pm0.10$\\
--- & --- & --- & --- & ---  & 5.5 & $1.833\pm0.12$\\
--- & --- & --- & --- & ---  & 6.5 & $1.656\pm0.12$\\
--- & --- & --- & --- & ---  & 7.5 & $1.535\pm0.12$\\
--- & --- & --- & --- & ---  & 8.5 & $1.370\pm0.11$\\
--- & --- & --- & --- & ---  & 9.0 & $1.259\pm0.08$\\
--- & --- & --- & --- & ---  & 9.5 & $1.100\pm0.07$\\
--- & --- & --- & --- & ---  & 10.5 & $1.006\pm0.08$\\
--- & --- & --- & --- & ---  & 11.0 & $0.884\pm0.06$\\
--- & --- & --- & --- & ---  & 11.5 & $0.825\pm0.06$\\
--- & --- & 23A-409 & 2023 May 26 & 964 & 2.5 & $0.737\pm0.05$ \\
--- & --- & --- & --- & ---  & 3.0 & $0.611\pm0.04$\\
--- & --- & --- & --- & ---  & 3.5 & $0.510\pm0.04$\\
--- & --- & --- & --- & ---  & 4.5 & $0.427\pm0.03$\\
--- & --- & --- & --- & ---  & 5.0 & $0.441\pm0.03$\\
--- & --- & --- & --- & ---  & 5.5 & $0.359\pm0.02$\\
--- & --- & --- & --- & ---  & 6.5 & $0.297\pm0.03$\\
--- & --- & --- & --- & ---  & 7.5 & $0.247\pm0.03$\\
--- & --- & --- & --- & ---  & 8.5 & $0.146\pm0.02$\\
--- & --- & --- & --- & ---  & 9.5 & $0.192\pm0.02$\\
--- & --- & --- & --- & ---  & 10.5 & $0.181\pm0.02$\\
--- & --- & --- & --- & ---  & 11.5 & $0.147\pm0.02$\\
--- & --- & --- & --- & ---  & 22.5 & $0.097\pm0.01$\\
\enddata
\tablecomments{$\delta t$ is measured relative to the first detection date. 
% (2019 Oct 20; \citealt{Hammerstein2023-ZTFsample}). 
%Flux densities are reported in mJy.
}
\end{deluxetable*}

\subsection{Sample Selection}

For the 53 radio TDEs collected, we select those suitable for CR analysis based on the following criteria: (1) at least two SEDs are available; 
and (2) at least two SEDs containing more than three data points. Applying these criteria, we yield a sample of 26 TDEs to constrain the CNM density profile index with the CR analysis. Here, we provide a brief overview of their basic properties.

\begin{enumerate}

  \item Sw J1644+57 was discovered on 2011 March 25 by the \emph{Swift} satellite, which is consistent with an inactivate galaxy at $z=0.354$ \citep{Bloom2011,Levan2011}. It is the first TDE with a relativistic jet, which brings a helpful tool to investigate the dormant SMBH at a more distant position. The radio observation started on 2011 March 29.36, about 3.9 days after the first $\gamma$-ray detection. The radio light curve of Sw J1644+57 shows a re-brightening about $30$ days after discovery \citep{Berger2012,Zauderer2013}, and it was well explained as a two-component jet model \citep{Wang2014,Liu2015,Teboul2023, Leiwh2026}. The overall CNM density profile is $k_{\rm eq}=1.5$ derived from equipartition analysis \citep{Eftekhari18}.

  % \item Swift J2058+05 was discovered by \emph{Swift} on 2011 May 17–20 at a redshift of $z=1.1853$. As the second jetted TDE after Swift J1644+57, it exhibits similar phenomenology \citep{Cenko2012-SwJ2058+05}. The available radio data from \citet{Cenko2012-SwJ2058+05} and \citet{Pasham2015-SwJ2058+05} are insufficient to construct a broadband SED, and thus its CNM density profile cannot be constrained.
  
  % \item Swift J1112-82 was discovered by the \emph{Swift} satellite in 2011 June 16-19, located in a galaxy with a redshift of $z=0.8901$ \citep{Brown2015-SwiftJ1112-82}. it is the third TDE with a relativistic jet. The scarcity of radio observations makes it impossible to carry out a robust analysis \citep{Brown2017-SwiftJ1112-82}.
  
  \item AT2022cmc was discovered on 2020 February 11 by ZTF, which locate at a distant galaxy with $z=1.19325$. It is the fourth on-axis jetted TDE, also the first jetted TDE discovered in optical \citep{Andreoni2022,Pasham2023,Teboul2023}. The first radio emission was obtained about $5$ days after its discovery \citep{Andreoni2022,Pasham2023}. Based on equipartition analysis and afterglow modeling, \citet{Matsumoto2023} roughly estimated the density profile of AT2022cmc is $k_{\rm eq}\simeq 1.5-2$, and generally consistent with a Bondi-like accretion. In our previous work, we have constrained the CNM density profile with both CRs and detailed \texttt{PyFRS}\footnote{\url{https://github.com/leiwh/PyFRS}} fit to the multi-band (optical, sub-millimeter, and radio) data of AT2022cmc, and we find $k_{\rm CR}\simeq 1.47$ and $k_{\rm PyFRS} \sim 1.68$ \citep{Zhou2024}.
  
  % \item EP250702a was first detected by Einstein Probe (EP) on 2025 July 2 at a redshift of $z= 1.036$ \citep{Cheng2025-EP250702a,Li2025-EP250702a,Gompertz2026-EP250702a_redshift}. Based on its multi-band properties, \citet{Li2025-EP250702a} proposed that EP250702a is highly likely to be a tidal disruption event of a white dwarf by an intermediate-mass black hole, and that it produced a relativistic jet. The first radio emission of EP250702a was detected by MeerKAT on 2025 July 4 (about 2.61\,days after the EP trigger) \citep{Bright2025-0702_first_radio}.
  
  % \item Arp 299-B AT1 was initially discovered in the near-infrared (NIR), coincident with the western nucleus B1 of the nearby ($z=0.010411$), luminous infrared galaxy Arp 299. Subsequent very long baseline interferometry (VLBI) radio observations resolved an expanding jet structure several years after the initial discovery \citep{Mattila2018-arp299}. However, the available radio data are limited, preventing the construction of a well-constrained spectral energy distribution (SED) for this source. Therefore, Arp 299-B AT1 is excluded from our analysis sample.
  
  \item ASASSN-14ae was triggered by ASAS-SN on 2014 January 25, located at the nuclear of a galaxy at $z=0.0436$ \citep{Holoien2014-14ae}. \citet{Cendes2024} first detected ASASSN-14ae at 2313 days at $6\rm GHz$ by VLA, and they also employed a late-time radio observation.
  
  \item ASASSN-14li was optically discovered by the All Sky Automated Survey for SuperNovae (ASAS-SN) on 2014 November 22, it is consistent with a nearby galaxy ($z=0.0206$) PGC 043234 \citep{Joes2014}. The initial radio data were obtained with the Arcminute Microkelvin Imager Large Array (AMI-LA) on 2014-12-23.08, about a month after the first optical discovery \citep{Velzen2016}. Since PGC 043234 had detected in radio before ASASSN-14li, probably due to a wake AGN, so we subtracted the radio emission from PGC 043234 like \citet{Alexander16}. Its radio emission also shows a standard synchrotron SED, the CNM profile from equipartition analysis is $k_{\rm eq}=2.5$ \citep{Alexander16}.

  \item ASASSN-15oi was discovered on 2015 August 14 at a redshift of $z=0.0484$ in optical \citep{Holoien2016}. It also detected in X-ray, UV, and radio. Radio observation started on 2015 August 22, $8$ days after the optical discovery, but without detections. Until 2016 February 12, a significant radio emission was detected \citep{Horesh2021-15oi}, and long-time radio observations of ASASSN-15oi show two delayed radio flares at $\sim 180$ days and $\sim 1400$ days after optical discovery \citep{Hajela2025-15oi}. \citet{Hajela2025-15oi} thought the first radio flare is due to stream-stream collisions.

  \item AT2018hyz/ASASSN-18zj was detected by ASAS-SN on 2018-11-06. It is consistent with a galaxy at $z=0.04573$ \citep{Hung2020-AT2018hyz, Short2020-AT2018hyz}. AT2018hyz showed a delayed and bright radio flare relative to its optical detection. The CNM density profile consistent with $k_{\rm eq}\sim1$ \citep{Cendes2022-AT2018hyz, Sfaradi2024-AT2018hyz,Teboul2023}. 
  %The analysis presented in \citet{Cendes2026_AT2018hyz_radio} indicates that AT2018hyz is consistent with a spherical outflow delayed by approximately 620 days, as well as with a highly off‑axis relativistic jet.
  
  % \item AT2018zr/PS18kh was first discovered on 2 March 2018 by Pan-STARRS1. The flare is related to a galaxy at a redshift of $z=0.071$ \citep{Holoien2019-18zr, Hung2019-18zr}. The radio observations by AMI and VLA on 28 March 2018 and 30 March 2018, which yielded no detections\citep{Velzen2019-18zr}. The first radio detections was obtained at $1713\rm days$ \citep{Cendes2024}. Currently, there is only one SED for its radio data, it is challenging for us to search the CNM density profile of AT2018zr.
  
  \item AT2019ahk/ASASSN-19bt was discovered on 2019 January 29.91 in the galaxy 2MASX J07001137-6602251, at a redshift of $z = 0.0262$. It is the first TDE flare detected by the Transiting Exoplanet Survey Satellite (TESS), and continued TESS observations show that the initial optical brighten happened on 2019 January 21.6 \citep{Holoien2019ASASSN-1bt}. Thus, we set the optical brighten time as the launch time of the outflow. The archival radio observations reveal a pre-disruption radio detection, which originates from the host galaxy. The initial radio detection related to ASASSN-19bt was obtained on 2019 March 3. We subtracted the host emission as described in \citet{Christy2024ASASSN-19bt}. It shows unusual late-time radio evolution, as mentioned in \citet{Christy2024ASASSN-19bt}, it might have a complex CNM environment.
  
  \item AT2019azh was first discovered by ASAS-SN on 2019 February 22, with a redshift of $z=0.022$ \citep{Brimacombe2019-19azh}. It was detected by ZTF on 2019 February 12 \citep{Velzen2019-19azh}. Long-time multiradio frequency observations of AT2019azh were presented by \citet{Goodwin2022-19azh}, they derived an irregular density profile by equipartiton. Combined with the $5.5\rm GHz$ light curve and the predicted decay rates for a Sedov–Taylor solution, \citet{Goodwin2022-19azh} obtained the CNM profile of AT2019azh of $k=2.5$. \citet{Sfaradi2022-19azh} offered a more detailed light curve of $15.5\rm GHz$, which is well fitted by two components of synchrotron emitting sources. In our analysis, we set the first detected time of ZTF as the onset of the outflow.
  
  \item AT2019dsg was discovered on 2019 April 09 as a nuclear transient at $z=0.0512$ by ZTF \citep{Nicholl2019-19dsg-z}. It was luminous in the X-ray, optical, UV and radio. On October 1, 2019, the IceCube Neutrino Observatory reported a neutrino detection $\sim 0.2\rm PeV$, it is the first TDE related to neutrino \citep{Stein2021}. The initial radio emission of AT2019dsg was obtained on 2019 May 21. % \citep{Cannizzaro2021-19dsg} thought a two-component jet and a changed CNM density profile could possible explain the radio emission.
  The CNM density profile was $k_{\rm eq}=1.7$ derived from equipartition analysis by using long-term radio observation \citep{Cendes2021-19dsg}.

  \item AT2019ehz was discovered on 2019 April 29 at a redshift of $z=0.074$ by Gaia \citep{Velzen2021}. The first radio emission was detected on 2021 June 11, about $775$ days after the Gaia detected time, it has two SEDs ($970$ days, $1262$ days) well fitted by synchrotron emission. The CNM density profile of AT2019ehz derived from \citet{Cendes2024} is $k_{\rm eq}>2.5$.
  
  \item AT2019eve was discovered by ZTF on 2019 May 5 at $z=0.0813$ \citep{Velzen2021}. The radio detection started on 2021 June 11, about $769$ days after the discovery time. It shows a late-time radio emission with three SEDs ($945$ days, $1102$ days, $1325$ days) consistent with synchrotron spectrum from a delayed outflow \citep{Cendes2024}. The CNM density profile obtained from equipartition is stepper than $2.5$ \citep{Cendes2024}.
  
  % \item AT2019qiz %QPE
  % is the most nearby TDE discovered to date. It was located at a redshift of $z=0.01513$, corresponds to a distance of $65.6\rm Mpc$. It was discovered on 2019-09-19 by ZTF \citep{Forster2019-AT2019qiz}, and the first detection was found from the Asteroid Terrestrial impact Last Alert System (ATLAS) on 2019-09-18 \citep{Nicholl2020-AT2019qiz}. \citet{Brien2019a-AT2019qiz,Brien2019b-AT2019qiz} had reported the first radio detections of AT2019qiz by ATCA. Recently, \citet{Anumarlapudi2024-radiosample} also given the lower-frequency observations of AT2019qiz. However, the radio data are still poor for us to employ an robust analysis.
  
  \item AT2019teq was discovered in a galaxy at $z = 0.087$ on 2019 October 20 in optical observations \citep{Hammerstein2023-ZTFsample}. An upper limit at radio frequencies was reported by VLASS at 351 days after the discovery. Later, \citet{Cendes2024} presented VLA detections at $1096\ \rm day$ and $1155\ \rm day$ post-discovery.

  \item AT2020opy %有能量注入，Fpeak随时间，不适用于我们的CR和Fpeak分析
  was first detected on 2020 July 8 by ZTF at $z=0.159$ \citep{Gezari2020-AT2020opy, Hammerstein2023-ZTFsample}. The radio emission of AT2020opy was initially obtained on 2020-10-6, about $\rm 90\, day$ after the ZTF detection. The CNM density profile around the SMBH obtained by equipartition is $k_{\rm eq}\sim 1.5-2.5$
  %, and the linear increase of the outflow energy means that there is a constant energy injected into the outflow 
  \citep{Goodwin2023-AT2020opy}.
  
  % \item AT2020pj was discovered by ZTF on 2020 January 2 at a redshift of $z=0.068$. The first radio emission data were reported by \citet{Cendes2024}; however, the existing data are insufficient to construct an SED, precluding further analysis.
  
  \item AT2020vwl was first detected on 2020 October by ZTF, located in a galaxy with a redshift of $z=0.0325$ \citep{Yao2023}. The first radio emission detected on 2021 February 23. Its long-time light curve reveal two radio flares\citep{Goodwin2023-AT2020vwl, Goodwin2025-AT2020vwl}. Although there was no previous significant AGN activity, there could be low-level host radio emission from low-luminosity AGN or star formation, so we subtracted host radio emission like \citet{Goodwin2023-AT2020vwl}. \citet{Goodwin2025-AT2020vwl} thought the second radio flare was due to an energy injection into the preexisting outflow or a second new outflow.
  
  % \item AT2020wey was discovered on 2020 October 8 by ZTF, and was classified as a TDE in a post-starburst galaxy at a redshift of $z=0.027$ \citep{Nordin2020-2020wey,Arcavi2020-2020wey}. The radio emission, detected 128 days post-discovery \citep{Cendes2024}, represents the only available radio data to date. However, this single epoch precludes a more detailed analysis.
  
  % \item AT2022dbl/ASASSN-22ci %repeating TDE
  % is a nearby TDE discovered by ASAS‑SN on 22 February 2022, with a redshift of 0.0284 \citep{Stanek2022-AT2022dbl, Arcavi2022-AT2022dbl}. Its optical emission faded to the host‑galaxy level within roughly a year of peak, but nearly two years after the initial outburst, ZTF detected a second brightening \citep{Yao2024-AT2022dbl}, the similarity between the two flares suggests that AT2022dbl is a multiple/repeating partial TDE \citep{Hinkle2024-AT2022dbl}. A faint radio signal of $32\pm7$ mJy at 15 GHz was detected with the VLA on 2022 February 26 \citep{Sfaradi2022-AT2022dbl} and no further radio activity has been reported to date.
  
  % \item AT2022dsb/eRASStJ154221.6-224012 was discovered by the extended ROentgen Survey with an Imaging Telescope Array (eROSITA; \citet{Predehl2021-eROSITA}) on 2022 February 17, consistent with the galaxy ESO 583-G004 at a redshift of $z=0.0235$. AT2022dsb was observed by the Australia Telescope Compact Array (ATCA) three times, only six radio data was obtained \citep{Malyali2024-22dsb}.

  \item AT2024tvd was discovered by ZTF on 25 August 2024 at a redshift of 0.044938, located $1.24\pm0.30$ arcsec from the host‑galaxy nucleus, identifying it as an offset TDE \citep{Sollerman2024-AT2024tvd, Yao2025-AT2024tvd}. \citet{Faris2024TNSCR-AT2024tvd} classified it as a TDE based on broad H and He II in the spectrum, central location in its host galaxy, and persistent UV emission. A radio counterpart was detected with VLA on 3 January 2025 ($\sim 131$ days after discovery), yielding a 10 GHz flux density of $0.60\pm0.06$ mJy \citep{Sfaradi2025TNSAN-AT2024tvd}. AMI subsequently measured $0.72\pm0.07$ mJy at 15.5 GHz on 13 January 2025 \citep{Horesh2025TNSAN-AT2024tvd}. Long‑term monitoring shows that its radio emission exhibits a double‑peaked evolution \citep{Sfaradi2025-AT2024tvd}.
  
  \item CNSS J0019+00 was identified in the Caltech–NRAO Stripe 82 Survey (CNSS) on 2015 March 21, it located in a nearby galaxy whose redshift is 0.018 \citep{Anderson2020-CNSS0019}. It is the first radio-discovered TDE, so we cannot set its discovery time as the onset of the outflow. It has no significant X-ray or optical emission. The radio spectra are well fit by a synchrotron spectrum, and the CNM profile of CNSS J0019+00 obtained from equipartition analysis is $k_{\rm eq}=2.5$ \citep{Anderson2020-CNSS0019}.

  \item AT2019avd was first detected by ZTF on 9 February 2019, located at the nuclear of a quiescent galaxy at a redshift of $z=0.029$ \citep{Malyali-19avd}. It was also detected in radio, IR, UV, and X-ray wavelengths. \citet{Wang2023-19avd} report the radio emission of AT2019avd from VLA and VLBA.%but there are only four radio detections
  
  \item AT2020wjw/eRASSt J234402.9-352640 was first discovered on 2020 November 28 by the eROSITA instrument, it coincides with the nuclear of the galaxy WISEA J234402.95-352641.8 at $z=0.1$ \citep{Homan2023}. The first radio emission was detected by \citet{Goodwin2024-J2344} on 2021 April 5 with ATCA, and they inferred the CNM density profile of J2344 is $k_{\rm eq}=1.4$ from equipartition modelling of the spectral properties. J2344 was detected in multiband that includes X-ray, UV, optical, infrared, and radio. The early X-ray and optical emission suggest J2344 likely a TDE in a turned-off AGN \citep{Homan2023}, therefore, we subtracted the radio emission from its host like \citet{Goodwin2024-J2344} before our analysis.

  \item eRASSt J011431–593654 ($z=0.16$) exhibited an X-ray flux peak detected by eROSITA on 2020 November 24. As reported in \citet{Goodwin2025-22eROSAT_sample}, this source was subsequently observed with the Australia Telescope Compact Array (ATCA) during two epochs separated by six months, with radio emission detected 1224\,days after the X-ray peak.
  
  % \item eRASSt J142140-295321 ($z=0.056$) exhibited an X-ray flux peak detected by eROSITA on 2020 January 30. As reported in \citet{Goodwin2025-22eROSAT_sample}, this source was subsequently observed with ATCA during two epochs separated by six months, with radio emission detected 1526\,days after the X-ray peak.
  
  % \item eRASSt J143624-174105 ($z=0.19$) exhibited an X-ray flux peak detected by eROSITA on 2022 February 18. As reported in \citet{Goodwin2025-22eROSAT_sample}, this source was subsequently observed with ATCA during two epochs separated by six months, with radio emission detected 777\,days after the X-ray peak.
   
  \item eRASSt J210858-562832 $z=0.043$ exhibited an X-ray flux peak detected by eROSITA on 2020 October 26. As reported in \citet{Goodwin2025-22eROSAT_sample}, this source was subsequently observed with ATCA during two epochs separated by six months. Through analysis of archival radio observations, they found a radio emission from the Australian Square Kilometre Array Pathfinder (ASKAP) Rapid ASKAP Continuum Survey (RACS) at 70.7\,days after the X-ray peak.

  % \item FIRST J1533+2727 is the second TDE discovered in radio. It was first detected by the Green Bank 300 foot telescope in 1986 and 1987 at $z=0.03243$. Interestingly, FIRST J1533+2727 was faded by a factor of 400 over $33\ \rm year$ at $5.5\ \rm GHz$ \citep{Ravi2022-FIRSTJ1533}. Unfortunately, the radio data are not good enough to obtain the CNM density profile with our CR analysis.
  
  \item IGR J12580+0134 was discovered by INTEGRAL as the tidal disruption of a super-Jupyter by a SMBH on 2011 January 2-11, associated to a nearby galaxy NGC 4845, at a redshift of $z=0.00411$ \citep{Nikolajuk2013}. The early time emission at $100$ to $857 \rm GHz$ based on Planck all-sky survey was shown in \citet{Yuan2016-IGR}, radio emission from IGR J12580+0134 has also been detected by JVLA at 1.57 and 6 GHz about one year after the X-ray peak \citep{Irwin2015-IGR}, and late-time radio emission detected by VLA and VLBA was given by \citet{Perlman2017-IGR}. IGR J12580+0134 is the first TDE with an off-axis relativistic jet \citep{Lei2016-IGR,Yuan2016-IGR}.
  %Radio emission from IGR J12580+0134 has been detected by JVLA at 1.57 and 6 GHz about one year after the X-ray peak \citep{Irwin2015-IGR}. The electron spectral index $p$ is determined through the in-band spectral measurements of the radio emission at 6 GHz (C-band), at which the emission is expected to be optically thin \citep{Irwin2015-IGR}. The C-band index of $\sim-0.587$ at T3 suggests an electron spectral index of $p \sim 2.17$. The low frequency (L-band at 1.57 GHz) emission is, however, in the optically thick regime, resulting in a turn-over of the spectra as shown in Figure 6. See Equation (24) for the analytical expression of the self-absorption frequency of the late stage jet evolution. The peak of the radio spectrum declines and shifts to lower frequencies with increasing time, which can be understood in terms of Equation (24). Our model can explain the general evolution trend of the L-band and C-band spectral indices.
  
  \item iPTF 16fnl was discovered on 2016-08-26 by the intermediate Palomar Transient Factory (iPTF) survey at $z=0.016328$ \citep{Gezari2016-16fnl}. iPTF 16fnl was detected in radio in the third epoch of the VLA observation at $\rm \sim 153\,day$ after the optical discovery, and showed a delayed radio flare \citep{Horesh2021-16fnl_radio, Cendes2024}.
  
  % \item OGLE17aaj was optically discovered by the 8.2 m Antu telescope of the Very Large Telescope (VLT) on 2017 January 2, it occurred in the nucleus of a weakly active galaxy at a redshift of $z=0.116$ \citep{Gromadzki2019-OGLE17aaj}. The first radio emission was detected at 1581 days after discovery \citep{Cendes2024}.
  
  \item PS16dtm/SN2016ezh was discovered at redshift of $z=0.0804$ on 12 August 2016 by Pan-STARRS1 \citep{Blanchard2017-PS16dtm, Jiang2017-PS16dtm,Petrushevska2023-PS16dtm}. It is a TDE happened in a pre-existing AGN. %\citep{Cendes2024}.

  \item WTP14adeqka was first detected by WISE on 20 June 2015, located at the redshift of $0.01895$. Its mid-infrared (MIR) light curve exhibits high luminosity, peaking at $\rm 10^{43}\, erg\,s^{-1}$ approximately , while showing no counterpart in optical or X-ray bands \citep{Masterson2024-WTP14adeqka}. \citet{Golay2025-WTP14adeqka} conducted a detailed radio study of WTP14adeqka at $4-10$ years post-disruption, thereby establishing it as the first mid-infrared TDE with detected delayed radio emission. We set the disruption time to 2015 April 19, following the IR light-curve modeling of \citet{Masterson2024-WTP14adeqka}. 
  
  \item XMMSL1 J074008.2-853927 (hereafter XMMSL1 J0740-85) was detected from a galaxy at a redshift of $z=0.0173$ by the XMM-Newton slew survey on 2014 April 1. It included both thermal and non-thermal emissions \citep{Saxton2017-XMMSL1}. The ATCA observation of XMMSL1 J0740-85 started on 2015 November 14, $592\rm d$ after the X-ray discovery \citep{Alexander2017-XMMSL1}. The CNM density profile of XMMSL1 J0740-85 is $k_{\rm eq}\sim1$ inferred from equipartition analysis \citep{Alexander2017-XMMSL1}.

  % \item eRASSt J045650.3-203750 %rptde
  % is a nuclear transient discovered in the eRASS2 all-sky survey between 2020 September 8 and 10, located in the nucleus of a quiescent galaxy at a redshift of $z = 0.077$ \citep{Liu2023-eRJ0456}. Long‑term X-ray and ultraviolet monitoring reveals recurrent outbursts with a characteristic recurrence timescale of approximately 223 days, indicating that the source is a repeating nuclear transient. Transient radio emission is also detected in eRASSt J045650.3-203750 \citep{Liu2024-eRJ0456}.

  % \item AT2019aalc %repeating TDE
  % was discovered by ZTF on January 22, 2019, and is located in a galaxy at redshift z = 0.0356. It exhibits multi-wavelength emission spanning radio, infrared, optical, ultraviolet, X-ray, and neutrino radiation. \citet{Veres2024-AT2019aalc-multband_radio} conducted multi-wavelength monitoring from radio to X-rays and identified two distinct optical flares. During the second optical flare, a notably bright UV counterpart and multiple X-ray flares were observed, accompanied by IR dust echo emission. According to the large-amplitude variability detected by VLASS, along with the high brightness temperature ($T_{\rm b} \gg 10^5 $ K) measured by EVN at 1.7 GHz, Therefore, AT2019aalc is interpreted as a TDE occurring within an AGN. Unfortunately, \citet{Veres2024-AT2019aalc-multband_radio} does not report any results from the radio data analysis.
  
  \item AT2020vdq %repeating TDE
  was first detected by ZTF on 4 October 2020, with a redshift of $z = 0.045$ \citep{Nordin2020-AT2020vdq}. Nearly 1000 days after the initial outburst, a new optical flare was detected near its position \citep{Charalampopoulos2023-AT2020vdq}, suggesting that AT2020vdq is a repeating partial TDE. \citet{Somalwar2025-AT2020vdq} reported that on 9 October 2021 approximately 383 days after the optical peak, VLASS detected a 3 GHz radio flux density of $1.48\pm 0.14$ mJy. Their multiwavelength analysis, including modeling of the radio emission, showed that the source can be described by a non-relativistic spherical synchrotron model. %Although Figure 3 in \citet{Somalwar2025-AT2020vdq} presents sufficient radio SEDs for a CR analysis, the data have not been provided, preventing us from performing CR analysis.
\end{enumerate}

% The information of these TDEs are summarized in Table \ref{tab:all_radio_tde}, and the radio light curves (around 5\,GHz) for these TDEs are plotted in Figure \ref{fig:radio TDEs}, except AT2019aalc, because its radio data are not published.
%These data can be available at \url{https://github.com/leiwh/RadioTDEs}.

\section{Constraining the CNM Profile with CR analysis} 
\label{sec:model}
The radio afterglow of TDEs can be interpreted as synchrotron emission originating from the forward shock (FS) driven by the interaction between the jet/outflow and the CNM. The flux density is typically characterized by a series of power-law segments \citep{Sari1998,Gao+2013,Zhang2018}, 
%$F_\nu \propto t^{-\alpha} \nu^{-\beta}$ .
\begin{equation}
    F_\nu \propto t^{-\alpha} \nu^{-\beta},
\label{eq:alphabeta}
\end{equation}
where $\alpha$ and $\beta$ are the temporal index and the spectral index, respectively. The relations between these two indices are called ``closure relations'' (CRs).

Within this framework, the type of CNM can be constrained using the CRs, following the same way as is commonly applied in gamma-ray burst (GRB) studies \citep{Gao+2013}. \citet{Gao+2013} derived the CRs for a constant density ($k=0$) and a wind type density profile ($k=2$). We obtain CRs for an arbitrary density profile, as described in several previous works \citep{Eerten2009, Fraija2021}.
%We extend the CRs derived by \citet{Gao+2013} for a constant density ($k=0$) and a wind type density profile ($k=2$) to the case of an arbitrary density profile as described in several previous works \citep{Eerten2009, Fraija2021}.

Generally, the jet evolution can be characterized by four phases. The first is a coasting phase, during which the Lorentz factor remains approximately constant, $\Gamma(t)\simeq \Gamma_0$. In the second phase, the shell starts to decelerate when the mass $m$ of the CNM swept by the FS is about $1/\Gamma_0$ of the rest mass in the ejecta $M_{\rm ej}$. The shell then approaches the \citet{BM1976} self-similar evolution. Subsequently, the ejecta enters the post-jet-break phase once the beaming angle $1/\Gamma$ becomes larger than the jet half-opening angle $\theta_{\rm j}$. Eventually, the blastwave transitions to the Newtonian deceleration phase after sweeping up CNM with a total rest-mass energy comparable to the initial energy of the ejecta, and the dynamics is governed by the well-known Sedov-Taylor solution. 

We consider a relativistic thin shell or a non-relativistic outflow expanding into the pre-existing CNM with density
\citep{Rees1992,Meszaros1997,Sari1998,Zhang2018,Huang2021}, 
\begin{equation}
    n(R)=n_{18} \left( \frac{R}{10^{18} \rm cm} \right)^{-k}=A R^{-k},
\label{eq:CNM_profile}
\end{equation}
where $n_{18}$ is the CNM density at $R=10^{18} \rm cm$. %We adopt an analytical description of the main properties of the evolution and emission of the FS from the jet/outflow-CNM interaction. 
During the interaction of jet/outflow and the CNM, electrons are accelerated at the shock front to a power-law distribution $N(\gamma_{\rm e}) \propto \gamma_{\rm e}^{-p}$. Under the assumption that two fractions $\epsilon_{\rm e}$ and $\epsilon_{\rm B}$ of the shock energy $e_2=4\Gamma^2 n m_{\rm p} c^2$ is deposited into electrons and magnetic field, respectively. This defines the minimum injected electron Lorentz factor \citep{Zhang2018,Sari1998},
\begin{equation}
\gamma_{\rm m}=\frac{p-2}{p-1} \epsilon_{\rm e} (\Gamma-1)\frac{m_{\rm p}}{m_{\rm e}},   
\end{equation}
and the comoving magnetic field
\begin{equation}
B=(32\pi m_{\rm p} \epsilon_{\rm B} n)^{1/2}\Gamma c.   
\end{equation}
where $m_{\rm e}$ and $m_{\rm p}$ are electron mass and proton mass. 
The synchrotron power and characteristic frequency emitted by an electron with Lorentz factor $\gamma_{\rm e}$ are given by \citep{Rybicki1979}
\begin{equation}
P(\gamma_{\rm e})\simeq \frac{4}{3} \sigma_{\rm T} c \Gamma^2 \gamma_{\rm e}^2 \frac{B^2}{8\pi},   
\end{equation}
\begin{equation}
\nu(\gamma_{\rm e}) \simeq \Gamma \gamma_{\rm e}^2 \frac{q_{\rm e} B}{2\pi m_{\rm e}c (1+z)},
\end{equation}
% where $P(\gamma_{\rm e})$ is expressed in the source frame, $\nu$ is measured in the observer frame, 
where $\sigma_{\rm T}$ is the Thomson cross-section, $q_{\rm e}$ is electron charge. The peak of the  spectra power occurs at $\nu(\gamma_{\rm e})$, and \citep{Sari1998,Gao+2013,Zhang2018}
\begin{equation}
P_{\nu,{\rm max}} \simeq \frac{P(\gamma_{\rm e})}{\nu(\gamma_{\rm e}) (1+z)}=\frac{m_{\rm e}c^2 \sigma_{\rm T}}{3q_{\rm e}} \Gamma B.   
\end{equation}
The critical electron Lorentz factor $\gamma_{\rm c}$ is defined by equating the electron's synchrotron cooling timescale to the observer-frame time $t$ \citep{Rybicki1979,Sari1998},
\begin{equation}
\gamma_{\rm c} = \frac{6\pi m_{\rm e}c}{\Gamma \sigma_{\rm T}B^2 t/(1+z)},
\end{equation}
% the electron distribution shape should be modified for $\gamma_{\rm e} >\gamma_{\rm c}$ when cooling due to synchrotron radiation becomes significant. 
Radiative cooling and continuous injection of accelerated electrons shape the electron distribution into a broken power law. The resulting synchrotron spectrum is then divided by three characteristic frequencies into multiple power-law segments \citep{Gao+2013,Zhang2018}. The first two characteristic frequencies $\nu_{\rm m}$ and $\nu_{\rm c}$ are determined by the two electron Lorentz factors $\gamma_{\rm \,m}$ and $\gamma_{\rm \, c}$, respectively. The third characteristic frequency is the self-absorption frequency $\nu_{\rm a}$, below which the synchrotron photons are self-absorbed. This frequency can be derived from both the optical depth method (i.e., where the optical depth for photon self-absorption is unity) and the blackbody method (i.e., where the synchrotron flux equals the blackbody radiation flux) \citep{Gao+2013,Zhang2018,Shen2009}. 
%This frequency can derived in two different ways. The first one is the optical depth method by the condition that the photon optical depth for self-absorption is unity ($\alpha_\nu(\nu_{\rm a})\Delta\sim 1$, where $\Delta$ is the characteristic width of the emission region) \citep{Rybicki1979}. Another way is the blackbody method by equating the synchrotron flux and the flux of a blackbody ($I_{\nu}^{\rm bb}(\nu_{\rm a})=I_{\nu}^{\rm syn}(\nu_{\rm a})=2kT\frac{\nu_{\rm a}^2}{c^2}$) \citep{Gao+2013,Zhang2018}. It can be proved that the two methods are equivalent to each other \citep{Shen2009}.
The maximum flux density is given by $F_{\nu,{\rm max}}=(1+z) N_{\rm e} P_{\nu,{\rm max}}/4\pi D^2$ \citep{Sari1998} , where $N_{\rm e}=\int 4\pi R^2 n dR$ is the total number of electrons in the shocked CNM and $D$ is the distance from the source.

\subsection{Closure Relations (CRs)}
Now we derive the temporal index $\alpha$ and the CRs $\alpha(\beta)$ in different outflow phases and for an arbitrary CNM density profile. Based on CRs, we then obtain the CNM density profile index $k_{\rm CR} (\alpha, \beta)$.

\label{sec:CR}
\subsubsection{Relativistic Jet in Coasting Phase}
%\subsubsection{Thin Shell Forward Shock in Coasting Phase}
The relativistic jet first undergoes a coasting phase, in which the dynamic evolution is given by
\begin{eqnarray}
\label{eq:BM_coasting_dyn}
\Gamma\simeq \Gamma_0, \\ \nonumber
 R(t)\simeq 2c\Gamma_0^2t .
\end{eqnarray}
We can then derive the scalings for FS spectra parameters in this phase as
\begin{equation}
\nu_{\rm m}\propto t^{-k/2},\ \ \nu_{\rm c} \propto t^{ \frac{3k-4}{2} }, \ \  F_{\nu, {\rm max} } \propto t^{\frac{6-3k}{2}} . 
\label{eq:sp_param_thinshell}
\end{equation}

For $\nu_{\rm a}<\nu_{\rm c}<\nu_{\rm m}$, we have $\nu_{\rm a} \propto t^{\frac{8-9k}{5}}$, and
\begin{eqnarray}
F_{\nu} = \left\lbrace
\begin{tabular}{l}
$F_{\nu, {\rm max}} \left(\frac{\nu_{\rm a}}{\nu_{\rm c}} \right)^{\frac{1}{3}}  \left(\frac{\nu}{\nu_{\rm a}} \right)^2  \propto t^{1+k} \nu^2  , \ \ \ \nu<\nu_{\rm a}$  \\
$F_{\nu, {\rm max}}\left( \frac{\nu}{\nu_{\rm c}}  \right)^{\frac{1}{3}} \propto t^{\frac{11-6k}{3}} \nu^{\frac{1}{3}} , \ \ \ \nu_{\rm a}<\nu<\nu_{\rm c}$ \\
$F_{\nu, {\rm max}}\left( \frac{\nu}{\nu_{\rm c}}  \right)^{-\frac{1}{2}} \propto t^{\frac{8-3k}{4}} \nu^{-\frac{1}{2}}  , \ \ \ \nu_{\rm c}<\nu<\nu_{\rm m}$  \\
$F_{\nu, {\rm max}}\left(\frac{\nu_{\rm m}}{\nu_{\rm c}} \right)^{-\frac{1}{2}}  \left(\frac{\nu}{\nu_{\rm m}} \right)^{-\frac{p}{2}} \propto t^{\frac{8-2k-kp}{4}} \nu^{-\frac{p}{2}} , \ \ \ \nu_{\rm m}<\nu$
\end{tabular} 
\right.
\end{eqnarray}
%where $\nu_{\rm a} \propto t^{\frac{8-9k}{5}}$.

For $\nu_{\rm a}<\nu_{\rm m}<\nu_{\rm c}$, we have $\nu_{\rm a} \propto t^{\frac{3-4k}{5}}$, and
\begin{eqnarray}
F_{\nu} =\left\lbrace
\begin{tabular}{l}
$F_{\nu, {\rm max}}\left(\frac{\nu_{\rm a}}{\nu_{\rm m}} \right)^{\frac{1}{3}}  \left(\frac{\nu}{\nu_{\rm a}} \right)^2  \propto t^2 \nu^2  , \ \ \ \nu<\nu_{\rm a}$  \\
$F_{\nu, {\rm max}}\left( \frac{\nu}{\nu_{\rm m}}  \right)^{\frac{1}{3}} \propto t^{\frac{9-4k}{3}} \nu^{\frac{1}{3}} , \ \ \ \nu_{\rm a}<\nu<\nu_{\rm m}$ \\
$F_{\nu, {\rm max}}\left( \frac{\nu}{\nu_{\rm m}}  \right)^{-\frac{p-1}{2}} \propto t^{\frac{12-5k-kp}{4}} \nu^{-\frac{p-1}{2}}  , \ \ \ \nu_{\rm m}<\nu<\nu_{\rm c}$  \\
$F_{\nu, {\rm max}}\left(\frac{\nu_{\rm c}}{\nu_{\rm m}} \right)^{-\frac{p-1}{2}}  \left(\frac{\nu}{\nu_{\rm c}} \right)^{-\frac{p}{2}} \propto t^{\frac{8-2k-kp}{4}} \nu^{-\frac{p}{2}} , \ \ \ \nu_{\rm c}<\nu$
\end{tabular} 
\right.
\end{eqnarray}
%where $\nu_{\rm a} \propto t^{\frac{3-4k}{5}}$.

For $\nu_{\rm m}<\nu_{\rm a}<\nu_{\rm c}$, we have $\nu_{\rm a} \propto t^{\frac{4-6k-kp}{2(p+4))}}$, and
\begin{eqnarray}
F_{\nu} =\left\lbrace
\begin{tabular}{l}
$F_{\nu, {\rm max}}\left(\frac{\nu_{\rm m}}{\nu_{\rm a}} \right)^{\frac{p+4}{2}}  \left(\frac{\nu}{\nu_{\rm m}} \right)^2   \propto t^2 \nu^2  , \ \ \ \nu<\nu_{\rm m}$  \\
$F_{\nu, {\rm max}}\left(\frac{\nu_{\rm a}}{\nu_{\rm m}} \right)^{-\frac{p-1}{2}}  \left(\frac{\nu}{\nu_{\rm a}} \right)^{\frac{5}{2}}  \propto t^{\frac{8+k}{4}} \nu^{\frac{5}{2}} , \ \ \ \nu_{\rm m}<\nu<\nu_{\rm a}$ \\
$F_{\nu, {\rm max}}\left(\frac{\nu}{\nu_{\rm m}} \right)^{-\frac{p-1}{2}}  \propto t^{\frac{12-5k-kp}{4}} \nu^{-\frac{p-1}{2}}  , \ \ \ \nu_{\rm a}<\nu<\nu_{\rm c}$  \\
$F_{\nu, {\rm max}}\left(\frac{\nu_{\rm c}}{\nu_{\rm m}} \right)^{-\frac{p-1}{2}} \left(\frac{\nu}{\nu_{\rm c}} \right)^{-\frac{p}{2}} \propto t^{\frac{8-2k-kp}{4}} \nu^{-\frac{p}{2}} , \ \ \ \nu_{\rm c}<\nu$
\end{tabular} 
\right.
\end{eqnarray}
%where $\nu_{\rm a} \propto t^{\frac{4-6k-kp}{2(p+4))}}$.
The $\alpha$, $\beta$ values, their CRs and the CNM density profile index $k_{\rm CR}$ obtained from CRs in this phase are
presented in Table \ref{tab:CR_coasting}.

%%%%%%%%%%%%%%%%%%%%%%%%%%%%%%%%%%%%%%%%
\begin{table}
\caption{ \textbf{Closure relations (CRs) for an arbitrary CNM profile $k$ in the coasting phase at different spectra regime}.} 
\label{tab:CR_coasting}
\centering
\fontsize{8}{11}\selectfont    %{字体尺寸}{行距}
%\begin{fourparttable}
\begin{tabular}{ccccc}
\toprule
\multirow{2}{*}{Spectra regime}&\multirow{2}{*}{$\beta$}&\multirow{2}{*}{$\alpha$}&\multirow{2}{*}{$\alpha(\beta)$}&\multirow{2}{*}{$k_{\rm CR}(\alpha,\beta)$} \\
 & & \\
\cmidrule(lr){1-5}%1-5列画横线
$\nu<\nu_a$&  -2& -(1+k) & $\frac{\beta}{2}(1+k)$ &   $\frac{2\alpha-\beta}{\beta}$ \\
$\nu_a<\nu<\nu_c$&  $-\frac{1}{3}$& $-\frac{11-6k}{3}$ &  $\beta(11-6k)$ &  $\frac{11\beta-\alpha}{6\beta}$\\
$\nu_c<\nu<\nu_m$&  $\frac{1}{2}$  & $\frac{3k-8}{4}$ & $\frac{\beta}{2} (3k-8)$ &$\frac{2\alpha+8\beta}{3\beta}$\\
$\nu_m<\nu$& $\frac{p}{2}$& $\frac{2k+kp-8}{4}$ & $\frac{k+k\beta -4}{2}$ &$\frac{2\alpha+4}{1+\beta}$\\
\cmidrule(lr){1-5}
$\nu<\nu_a$& $-2$&  -2 &  $\beta$&  $-$\\
$\nu_a<\nu<\nu_m$& $-\frac{1}{3}$& $-\frac{9-4k}{3}$ & $\beta(9-4k)$ & $\frac{9\beta-\alpha}{4\beta}$\\
$\nu_m<\nu<\nu_c$& $\frac{p-1}{2}$& $\frac{5k+kp-12}{4}$ & $\frac{3k+k\beta-6}{2}$ &     $\frac{2\alpha+6}{3+\beta}$\\
$\nu_c<\nu$& $\frac{p}{2}$& $\frac{2k+kp-8}{4}$ & $\frac{k+k\beta-4}{2}$ & $\frac{2\alpha+4}{1+\beta}$\\
\cmidrule(lr){1-5}
$\nu<\nu_m$& -2& -2 & $\beta$ & $-$\\
$\nu_m<\nu<\nu_a$& $-\frac{5}{2}$& $-\frac{8+k}{4}$ & $ \frac{\beta(8+k)}{10}$ &   $\frac{10\alpha-8\beta}{\beta}$\\
$\nu_a<\nu<\nu_c$& $\frac{p-1}{2}$& $\frac{5k+kp-12}{4}$ & $\frac{3k+k\beta-6}{2}$ &     $\frac{2\alpha+6}{3+\beta}$\\
$\nu_c<\nu$& $\frac{p}{2}$&$\frac{2k+kp-8}{4}$ & $\frac{k+k\beta-4}{2}$ & $\frac{2\alpha+4}{1+\beta}$\\
%\cmidrule(lr){1-4}

\bottomrule
\end{tabular}\vspace{0cm}
%\end{fourparttable}
\end{table}
%%%%%%%%%%%%%%%%%%%%%%%%%%%%%%%%%%%%%%%%

\subsubsection{Relativistic Jet in Deceleration Phase}
%Blandford-McKee deceleration jet
For a relativistic jet, it will approach the \citet{BM1976} self-similar evolution when it is decelerated. The dynamic evolution of the blastwave for arbitrary $k$ density profile is given by
\begin{eqnarray}
\label{eq:BM}
\Gamma(t)\simeq \left(\frac{(17-4k)E_{\rm K,iso}}{4^{5-k} (4-k)^{3-k} \pi A m_{\rm p} c^{5-k} t^{3-k} }  \right)^{\frac{1}{2(4-k)}}, \\ \nonumber
 R(t)\simeq \left(\frac{(17-4k)(4-k) E_{\rm K,iso} t}{4 \pi A m_{\rm p} c}  \right)^{\frac{1}{(4-k)}} ,
\end{eqnarray}
% \begin{equation}
% \Gamma(t)\simeq \left(\frac{(17-4k)E_{\rm K,iso}}{4^{5-k} (4-k)^{3-k} \pi A m_{\rm p} c^{5-k} t^{3-k} }  \right)^{\frac{1}{2(4-k)}}, \ \ \  R(t)\simeq \left(\frac{(17-4k)(4-k) E_{\rm K,iso} t}{4 \pi A m_{\rm p} c}  \right)^{\frac{1}{(4-k)}} ,
% \label{eq:BM}
% \end{equation}
\noindent where $m_{\rm p}$ is proton mass. From Equation (\ref{eq:BM}), $\Gamma \propto t^{-\frac{(3-k)}{2(4-k)}}$ and $R\propto t^{\frac{1}{4-k}}$, one has the scalings for the FS spectra parameters in this phase as 
\begin{equation}
\label{eq:sp_param_BM}
\nu_{\rm m}\propto t^{-3/2},\ \ \nu_{\rm c} \propto t^{ \frac{4-3k}{2(k-4)} }, \ \  F_{\nu, {\rm max} } \propto t^{\frac{k}{2(k-4)}} . 
\end{equation}

%\R{$\nu_{\rm a}<\nu_{\rm c}<\nu_{\rm m}$ regime}
For $\nu_{\rm a}<\nu_{\rm c}<\nu_{\rm m}$, we have $\nu_{\rm a} \propto t^{\frac{3k+10}{5(k-4)}}$, and
\begin{eqnarray}
F_{\nu} \propto \left\lbrace
\begin{tabular}{l}
$ t^{\frac{4}{4-k}} \nu^2  , \ \ \ \nu<\nu_{\rm a}$  \\
$ t^{\frac{3k-2}{3(k-4)}} \nu^{\frac{1}{3}} , \ \ \ \nu_{\rm a}<\nu<\nu_{\rm c}$ \\
$ t^{-\frac{1}{4}} \nu^{-\frac{1}{2}}  , \ \ \ \nu_{\rm c}<\nu<\nu_{\rm m}$  \\
$t^{\frac{2-3p}{4}} \nu^{-\frac{p}{2}} , \ \ \ \nu_{\rm m}<\nu$
\end{tabular} 
\right.
\end{eqnarray}

For $\nu_{\rm a}<\nu_{\rm m}<\nu_{\rm c}$, we have $\nu_{\rm a} \propto t^{-\frac{3k}{5(4-k)}}$, and
\begin{eqnarray}
F_{\nu} \propto \left\lbrace
\begin{tabular}{l}
$ t^{\frac{2}{4-k}} \nu^2  , \ \ \ \nu<\nu_{\rm a}$  \\
$ t^{\frac{2-k}{4-k}} \nu^{\frac{1}{3}} , \ \ \ \nu_{\rm a}<\nu<\nu_{\rm m}$ \\
$ t^{-\frac{12p-12+5k-3kp}{16-4k}} \nu^{-\frac{p-1}{2}}  , \ \ \ \nu_{\rm m}<\nu<\nu_{\rm c}$  \\
$ t^{-\frac{3p-2}{4}} \nu^{-\frac{p}{2}} , \ \ \ \nu_{\rm c}<\nu$
\end{tabular} 
\right.
\end{eqnarray}

For $\nu_{\rm m}<\nu_{\rm a}<\nu_{\rm c}$, we have $\nu_{\rm a} \propto t^{\frac{2k-3kp+8+12p}{2(p+4)(k-4)}}$, and
\begin{eqnarray}
F_{\nu} \propto \left\lbrace
\begin{tabular}{l}
% $\left(\frac{\nu_{\rm a}}{\nu_{\rm m}} \right)^{-\frac{p-1}{2}}  \left(\frac{\nu_{\rm m}}{\nu_{\rm a}} \right)^{\frac{5}{2}}  \left(\frac{\nu}{\nu_{\rm m}} \right)^2 = \left(\frac{\nu_{\rm m}}{\nu_{\rm a}} \right)^{\frac{p+4}{2}}  \left(\frac{\nu}{\nu_{\rm m}} \right)^2   \propto t^{\frac{2}{4-k}} \nu^2  , \ \ \ \nu<\nu_{\rm m}$  \\
$ t^{\frac{2}{4-k}} \nu^2  , \ \ \ \nu<\nu_{\rm m}$  \\
$ t^{\frac{20-3k}{4(4-k)}} \nu^{\frac{5}{2}} , \ \ \ \nu_{\rm m}<\nu<\nu_{\rm a}$ \\
$ t^{\frac{12+3kp-5k-12p}{4(4-k)}} \nu^{-\frac{p-1}{2}}  , \ \ \ \nu_{\rm a}<\nu<\nu_{\rm c}$  \\
$ t^{\frac{2-3p}{4}} \nu^{-\frac{p}{2}} , \ \ \ \nu_{\rm c}<\nu$
\end{tabular} 
\right.
\end{eqnarray}

In Table \ref{tab:CR_BM}, we also present the $\alpha$, $\beta$ values, their CRs and the CNM density profile index $k_{\rm CR}$.

The results are consistent with the results of \citet[see Tables 1 and 2 therein]{Eerten2009}, but they do not include the self-absorption frequency $\nu_{\rm a}$.

%%%%%%%%%%%%%%%%%%%%%%%%%%%%%%%%%%%%%%%%
\begin{table}
\caption{ \textbf{Closure relations (CRs) for an arbitrary CNM profile $k$ in the BM deceleration phase at different spectra regime}.} 
\label{tab:CR_BM}
\centering
\fontsize{8}{11}\selectfont    %{字体尺寸}{行距}
%\begin{fourparttable}
\begin{tabular}{ccccc}
\toprule
\multirow{2}{*}{Spectra regime}&\multirow{2}{*}{$\beta$}&\multirow{2}{*}{$\alpha$}&\multirow{2}{*}{$\alpha(\beta)$}&\multirow{2}{*}{$k_{\rm CR}(\alpha,\beta)$} \\
 & & \\
\cmidrule(lr){1-5}%1-5列画横线
$\nu<\nu_a$&  -2& $-\frac{4}{4-k}$ & $\frac{2\beta}{4-k}$ &  $\frac{4\alpha-2\beta}{\alpha}$ \\
$\nu_a<\nu<\nu_c$&  $-\frac{1}{3}$&  $-\frac{3k-2}{3(k-4)}$ & $ \frac{\beta(3k-2)}{k-4}$ &  $\frac{4\alpha-2\beta}{\alpha-3\beta}$\\
$\nu_c<\nu<\nu_m$&  $\frac{1}{2}$  &  $\frac{1}{4}$ & $\frac{\beta}{2}$  &$-$\\
$\nu_m<\nu$& $\frac{p}{2}$ & $\frac{3p-2}{4}$  &  $\frac{3\beta-1}{2}$ &$-$\\
\cmidrule(lr){1-5}
$\nu<\nu_a$& $-2$& $-\frac{2}{4-k}$ & $\frac{\beta}{4-k}$ &  $\frac{4\alpha-\beta}{\alpha}$\\
$\nu_a<\nu<\nu_m$& $-\frac{1}{3}$& $-\frac{k-2}{k-4}$ & $ \frac{3\beta(k-2)}{k-4}$ &  $\frac{4\alpha-6\beta}{\alpha-3\beta}$\\
$\nu_m<\nu<\nu_c$& $\frac{p-1}{2}$& $\frac{12p-12+5k-3kp}{16-4k}$ & $\frac{12\beta+k-3k\beta}{8-2k}$ &  $\frac{8\alpha-12\beta}{2\alpha-3\beta+1}$\\
$\nu_c<\nu$& $\frac{p}{2}$ & $\frac{3p-2}{4}$ & $\frac{3\beta-1}{2}$ & $-$ \\
\cmidrule(lr){1-5}
$\nu<\nu_m$& -2& $-\frac{2}{4-k}$ & $\frac{\beta}{4-k}$ &  $\frac{4\alpha-\beta}{\alpha}$\\
$\nu_m<\nu<\nu_a$& $-\frac{5}{2}$&  $-\frac{20-3k}{4(4-k)}$  & $\frac{\beta(20-3k)}{10(4-k)}$ &$\frac{40\alpha-20\beta}{10\alpha-3\beta}$\\
$\nu_a<\nu<\nu_c$& $\frac{p-1}{2}$& $\frac{12p-12+5k-3kp}{16-4k}$ & $\frac{12\beta+k-3k\beta}{8-2k}$ &  $\frac{8\alpha-12\beta}{2\alpha-3\beta+1}$\\
$\nu_c<\nu$& $\frac{p}{2}$&$\frac{3p-2}{4}$ & $\frac{3\beta-1}{2}$ & $-$ \\

%\cmidrule(lr){1-4}

\bottomrule
\end{tabular}\vspace{0cm}
%\end{fourparttable}
\end{table}

\subsubsection{Non-relativistic Outflow in Coasting Phase}
For a non-relativistic outflow, the magnetic field is given by 
\begin{equation}
    B=\left(4\pi m_{\rm p}\epsilon_{\rm B} n\right)^{1/2}v,
\end{equation}
\noindent where $v$ is the velocity of the blastwave. The minimum electron Lorentz factor is
\begin{equation}
    \gamma_{\rm m}=\epsilon_{\rm e}\frac{p-2}{p-1}\frac{m_{\rm p}}{m_{\rm e}}\frac{v^2}{c^2}\frac{1}{2}.
\end{equation}
 When the outflow in the coasting phase, the dynamic evolution can be described by
 \begin{eqnarray}
\label{eq:newtonian_coasting_dyn}
v\simeq v_0, \\ \nonumber
 R(t)\simeq vt.
\end{eqnarray}

From Equation (\ref{eq:newtonian_coasting_dyn}), $R \propto t$ and $v$ are independent of time, one has the scalings for the FS spectra parameters in this phase as $\nu_{\rm m}\propto t^{-k/2}$, $\nu_{\rm c} \propto t^{ \frac{3k-4}{2}}$, and $F_{\nu, {\rm max}} \propto t^{\frac{6-3k}{2}}$, similar to the case of a relativistic jet in coasting phase. Therefore, a relativistic outflow and a non-relativistic outflow share the same CRs and also $k_{\rm CR}(\alpha, \beta)$ during the coasting phase (See Table \ref{tab:CR_coasting}).

\subsubsection{Non-relativistic Outflow in Deceleration Phase}
If the outflow is Newtonian, the dynamic evolution can be described by the Sedov-Taylor (ST) solution in the deceleration phase,
\begin{eqnarray}
\label{eq:ST} 
    R(t)\simeq \left( \frac{5-k}{2}\right)^{\frac{2}{5-k}} \left[ \frac{\left(3-k\right)E_{\rm k,iso}}{2\pi Am_{\rm p}}\right]^\frac{1}{5-k} t^\frac{2}{5-k}, \\ \nonumber
    v(t)\simeq \left( \frac{5-k}{2}\right)^{\frac{k-3}{5-k}}\left[ \frac{\left(3-k\right)E_{\rm k,iso}}{2\pi Am_{\rm p}}\right]^\frac{1}{5-k} t^\frac{k-3}{5-k}.  
\end{eqnarray}
% \begin{equation}
%     R(t)\simeq \left( \frac{5-k}{2}\right)^{\frac{2}{5-k}} \left[ \frac{\left(3-k\right)E}{2\pi Am_p}\right]^\frac{1}{5-k} t^\frac{2}{5-k}, \ \ \ v(t)\simeq \left( \frac{5-k}{2}\right)^{\frac{k-3}{5-k}}\left[ \frac{\left(3-k\right)E}{2\pi Am_p}\right]^\frac{1}{5-k} t^\frac{k-3}{5-k},
% \label{eq:ST}
% \end{equation}

From Equation (\ref{eq:ST}), $R \propto t^{\frac{1}{5-k}}$ and $v \propto t^{\frac{k-3}{5-k}}$, one has the scalings for the FS spectra parameters in this phase as
\begin{equation}
\nu_{\rm m}\propto t^{\frac{4k-15}{5-k}},\ \ \nu_{\rm c} \propto t^{ \frac{2k-1}{5-k} }, \ \  F_{\nu, {\rm max} } \propto t^{\frac{3-2k}{5-k}} .  
\label{eq:fpeak_NT}
\end{equation}

For $\nu_{\rm a}<\nu_{\rm m}<\nu_{\rm c}$, we have $\nu_{\rm a} \propto t^{\frac{30-16k}{5(5-k)}}$, and
\begin{eqnarray}
F_{\nu}\propto  \left\lbrace
\begin{tabular}{l}
$ t^{\frac{2(k-1)}{5-k}} \nu^2  , \ \ \ \nu<\nu_{\rm a}$ \\
$ t^{\frac{24-10k}{3(5-k)}} \nu^{\frac{1}{3}} , \ \ \ \nu_{\rm a}<\nu<\nu_{\rm m}$ \\
$ t^{\frac{21-8k+4kp-15p}{2(5-k)}} \nu^{-\frac{p-1}{2}}  , \ \ \ \nu_{\rm m}<\nu<\nu_{\rm c}$  \\
$ t^{\frac{20-6k+4kp-15p}{2(5-k)}} \nu^{-\frac{p}{2}} , \ \ \ \ \nu>\nu_{\rm c}$
\end{tabular} 
\right.
\end{eqnarray}

For $\nu_{\rm m}<\nu_{\rm a}<\nu_{\rm c}$, we have $\nu_{\rm a} \propto t^{\frac{10-8k+4kp-15p}{(5-k)(p+4)}}$, and
\begin{eqnarray}
F_{\nu} \propto \left\lbrace
\begin{tabular}{l}
$ t^{\frac{2(k-1)}{5-k}} \nu^2  , \ \ \ \nu<\nu_{\rm m}$  \\
$ t^{\frac{11}{2(5-k)}} \nu^{\frac{5}{2}} , \ \ \ \nu_{\rm m}<\nu<\nu_{\rm a}$ \\
$ t^{\frac{21-8k+4kp-15p}{2(5-k)}} \nu^{-\frac{p-1}{2}}  , \ \ \ \nu_{\rm a}<\nu<\nu_{\rm c}$  \\
$ t^{\frac{20-6k+4kp-15p}{2(5-k)}} \nu^{-\frac{p}{2}} , \ \ \ \nu_{\rm c}<\nu$
\end{tabular} 
\right.
\end{eqnarray}

For $k=0$ and 2, our results return to those given by \citet{Gao+2013}.

We summarize the $\alpha$, $\beta$ values, their CRs and the CNM density profile index $k_{\rm CR}$ in Table \ref{tab:CR_ST}. Therefore, the measurements of $\alpha$ and $\beta$, can be used to estimate the CNM density profile ($k$). 

%%%%%%%%%%%%%%%%%%%%%%%%%%%%%%%%%%%%%%%%
\begin{table}
\caption{ Closure relations (CRs) for an arbitrary CNM profile $k$ in the ST deceleration phase at different spectra regime.} 
\label{tab:CR_ST}
\centering
\fontsize{8}{11}\selectfont    %{字体尺寸}{行距}
%\begin{fourparttable}
\begin{tabular}{ccccc}
\toprule
\multirow{2}{*}{Spectra regime}&\multirow{2}{*}{$\beta$}&\multirow{2}{*}{$\alpha$}&\multirow{2}{*}{$\alpha(\beta)$}&\multirow{2}{*}{$k_{\rm CR}(\alpha,\beta)$} \\
 & & \\
\cmidrule(lr){1-5}%1-5列画横线
$\nu<\nu_a$& -2& $-\frac{2(k-1)}{5-k}$  & $ \frac{\beta(k-1)}{5-k}$ &  $\frac{5\alpha+\beta}{\alpha+\beta}$\\
$\nu_a<\nu<\nu_m$& $-\frac{1}{3}$& $-\frac{24-10k}{3(5-k)}$  & $ \frac{\beta(24-10k)}{5-k}$ & $\frac{5\alpha-24\beta}{\alpha-10\beta}$\\
$\nu_m<\nu<\nu_c$& $\frac{p-1}{2}$&  $\frac{15p-4kp+8k-21}{2(5-k)}$  &  $\frac{15\beta-3+2k-4k\beta}{5-k}$  &  $\frac{5\alpha-15\beta+3}{\alpha-4\beta+2}$\\
$\nu_c<\nu$& $\frac{p}{2}$& $\frac{15p-4kp+6k-20}{2(5-k)}$ & $\frac{15\beta-10+3k-4k\beta}{5-k}$ &   $\frac{5\alpha-15\beta+10}{\alpha-4\beta+3}$\\
\cmidrule(lr){1-5}
$\nu<\nu_m$& -2&  $-\frac{2(k-1)}{5-k}$  & $ \frac{\beta(k-1)}{5-k}$ &  $\frac{5\alpha+\beta}{\alpha+\beta}$\\
$\nu_m<\nu<\nu_a$& $-\frac{5}{2}$& $-\frac{11}{2(5-k)}$ & $ \frac{11\beta}{5(5-k)}$ &$\frac{25\alpha-11\beta}{5\alpha}$\\
$\nu_a<\nu<\nu_c$& $\frac{p-1}{2}$&  $\frac{15p-4kp+8k-21}{2(5-k)}$  &  $\frac{15\beta-3+2k-4k\beta}{5-k}$  &  $\frac{5\alpha-15\beta+3}{\alpha-4\beta+2}$\\
$\nu_c<\nu$& $\frac{p}{2}$& $\frac{15p-4kp+6k-20}{2(5-k)}$ & $\frac{15\beta-10+3k-4k\beta}{5-k}$ &   $\frac{5\alpha-15\beta+10}{\alpha-4\beta+3}$\\

%\cmidrule(lr){1-4}

\bottomrule
\end{tabular}\vspace{0cm}
%\end{fourparttable}
\end{table}

\subsubsection{Deep-Newtonian Phase}
When the outflow decelerates below a critical velocity $v_{\rm DN}$ ($v_{\rm DN}=(\frac{8m_{\rm e}}{m_{\rm p}\bar{\epsilon}_{\rm e}})^{1/2}c$, where $\bar{\epsilon}_{\rm e}=4\epsilon_{\rm e}(p-2)/(p-1)$), the Lorentz factor is fixed to $\gamma_{\rm m} = 2$. This regime is called the deep-Newtonian phase\citep{Huang2003,Sironi2013,Matsumoto2021-DN,Matsumoto2022}. During the deep-Newtonian phase, only a fraction of particles accelerate into the nonthermal distribution, the fraction is given as \citet{Goodwin2025-AT2020vwl},
\begin{equation}
f_{\rm NT} = 
\begin{cases} 
x \propto t^{2(k-3)/(5-k)}, & 2 < p < 3\\ 
x^{(p-1)/2} \propto t^{(p-1)(k-3)/(5-k)},& p \gtrsim 3 
\end{cases}
\end{equation}
where $x=\epsilon_{\rm e} \frac{(v/c)^2}{2})\frac{m_{\rm p}}{m_{\rm e}}$.
Therefore, we have the scalings for the FS spectra parameters in this phase and for $\nu_{\rm m}<\nu_{\rm a}<\nu_{\rm c}$ as
\begin{equation}
\nu_{\rm m}\propto t^{\frac{-3}{5-k}},\ \ \nu_{\rm c} \propto t^{ \frac{2k-1}{5-k} }, \ \ \nu_{\rm a}\propto t^{\frac{-14-3p}{(5-k)(p+4)}},  
\label{eq:fpeak_NT}
\end{equation}
and
\begin{equation}
F_{\nu, \text{max}} \propto 
\begin{cases} 
t^{-3/(5-k)}, & 2 < p < 3 \\ 
t^{(6-3k+pk-3p)/(5-k)}, & p \gtrsim 3
\end{cases}.
\end{equation}
We then obtain,
\begin{eqnarray}
F_{\nu} \propto  \left\lbrace
\begin{tabular}{l}
$ t^{\frac{4}{5-k}} \nu^2  , \ \ \ \nu<\nu_{\rm m}$  \\
$ t^{\frac{11}{2(5-k)}} \nu^{\frac{5}{2}} , \ \ \ \nu_{\rm m}<\nu<\nu_{\rm a}$ \\
$ t^{\frac{-3-3p}{2(5-k)}} \nu^{-\frac{p-1}{2}}  , \ \ \ \nu_{\rm a}<\nu<\nu_{\rm c}$  \\
$ t^{\frac{2k-4-3p}{2(5-k)}} \nu^{-\frac{p}{2}} , \ \ \ \nu_{\rm c}<\nu$
\end{tabular} 
\right.
\end{eqnarray}
for $2<p<3$, and,
\begin{eqnarray}
F_{\nu} \propto \left\lbrace
\begin{tabular}{l}
$ t^{\frac{13-3k+pk-3p}{5-k}} \nu^2  , \ \ \ \nu<\nu_{\rm m}$  \\
$ t^{\frac{2pk-6p-6k+29}{2(5-k)}} \nu^{\frac{5}{2}} , \ \ \ \nu_{\rm m}<\nu<\nu_{\rm a}$ \\
$ t^{\frac{2pk-9p-6k+15}{2(5-k)}} \nu^{-\frac{p-1}{2}}  , \ \ \ \nu_{\rm a}<\nu<\nu_{\rm c}$  \\
$ t^{\frac{2pk-9p-4k+14}{2(5-k)}} \nu^{-\frac{p}{2}} , \ \ \ \nu_{\rm c}<\nu$
\end{tabular} 
\right.
\end{eqnarray}
for $p\gtrsim3$. In the same way, we summarize $\alpha$, $\beta$, their CRs and the CNM density profile index $k_{\rm CR}$ in Table \ref{tab:DN_exp}.

If the outflow enters the deep-Newtonian regime in the coasting phase, the fraction $f_{\rm NT}$ will be time-independent, since it is a function of the flow velocity $v$. Therefore, the CRs and also $k_{\rm CR}(\alpha,\beta)$ will be the same as the coasting phase for a relativistic outflow (see Table \ref{tab:CR_coasting}).

%%%%%%%%%%%%%%%%%%%%%%%%%%%%%%%%%%%%%%%%%%%%%%%%%%
\begin{table}
\caption{CRs for an arbitrary CNM profile $k$ in deep-Newtonian phase.} 
\label{tab:DN_exp}
\centering
\fontsize{8}{11}\selectfont    %{字体尺寸}{行距}
%\begin{threeparttable}
\begin{tabular}{ccccc}
\toprule
\multirow{2}{*}{Spectra regime}&\multirow{2}{*}{$\beta$}&\multirow{2}{*}{$\alpha$}&\multirow{2}{*}{$\alpha(\beta)$}&\multirow{2}{*}{$k_{\rm CR}(\alpha,\beta)$} \\
 & & \\
\cmidrule(lr){1-5}%1-5列画横线
$2<p<3$ & \\
$\nu<\nu_m$&  $-2$& $-\frac{4}{5-k}$ & $\frac{2\beta}{5-k}$ &   $\frac{5\alpha-2\beta}{\alpha}$ \\
$\nu_m<\nu<\nu_a$&  $-\frac{5}{2}$& $-\frac{11}{2(5-k)}$  & $\frac{11\beta}{5(5-k)}$ & $\frac{25\alpha-11\beta}{5\alpha}$\\
$\nu_a<\nu<\nu_c$&  $\frac{p-1}{2}$ & $\frac{3(1+p)}{2(5-k)}$ & $\frac{3(1+\beta)}{5-k}$   &$\frac{5\alpha-3\beta-3}{\alpha}$\\
$\nu_c<\nu$& $\frac{p}{2}$ &  $\frac{3p+4-2k}{2(5-k)}$ & $\frac{3\beta+2-k}{5-k}$ &$\frac{5\alpha-3\beta-2}{\alpha-1}$\\
\cmidrule(lr){1-5}
$p\gtrsim3$ &\\
$\nu<\nu_m$&  $-2$& $-\frac{kp-3p-3k+13}{5-k}$ & $\frac{\beta(kp-3p-3k+13)}{2(5-k)}$ &  $\frac{10\alpha-13\beta+3p\beta}{2\alpha-3\beta+p\beta}$ \\
$\nu_m<\nu<\nu_a$&  $-\frac{5}{2}$& $-\frac{2kp-6p-6k+29}{2(5-k)}$ & $\frac{\beta(2kp-6p-6k+29)}{5(5-k)}$ & $\frac{25\alpha-29\beta+6p\beta}{5\alpha-6\beta+2p\beta}$\\
$\nu_a<\nu<\nu_c$&  $\frac{p-1}{2}$   & $\frac{9p+6k-2kp-15}{2(5-k)}$ &  $\frac{9\beta+2k-2k\beta-3}{5-k}$ &$\frac{5\alpha-9\beta+3}{\alpha-2\beta+2}$\\
$\nu_c<\nu$& $\frac{p}{2}$ &  $\frac{9p+4k-2kp-14}{2(5-k)}$ & $\frac{9\beta+2k-2k\beta-7}{5-k}$ &$\frac{5\alpha-9\beta+7}{\alpha-2\beta+2}$\\
\bottomrule
\end{tabular}\vspace{0cm}
%\end{threeparttable}
\end{table}

\begin{figure*}[htp]
%\begin{figure}[htp]
\center
\includegraphics[width=1.0\textwidth]{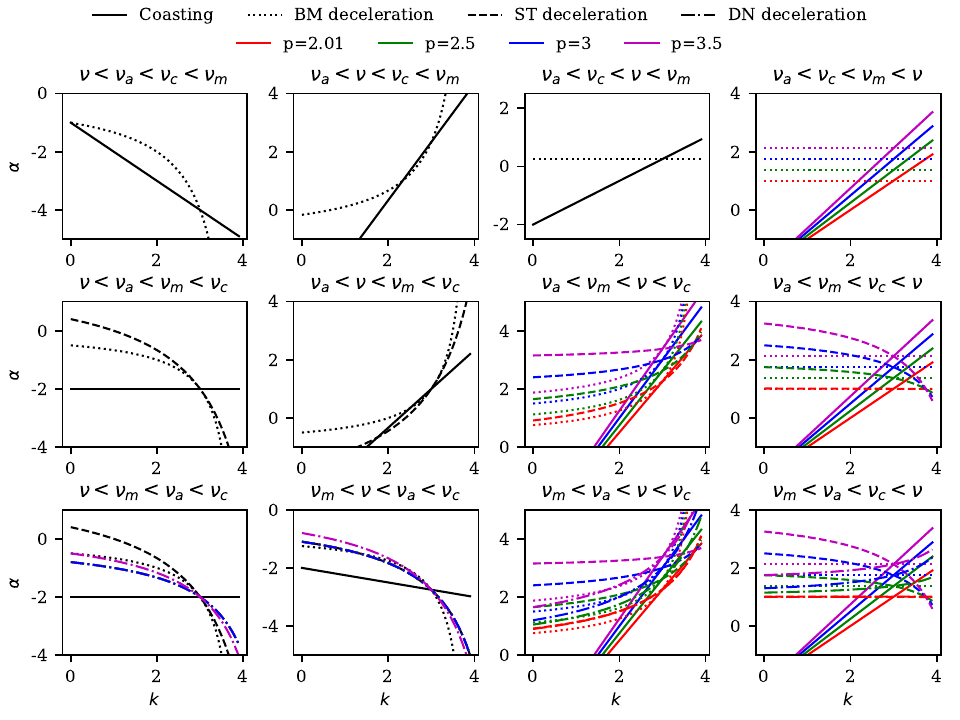}
\caption{The relation between temporal index $\alpha$ and the CNM density profile $k$ for $\nu_a<\nu_c<\nu_m$ (top), $\nu_{\rm a}<\nu_{\rm m}<\nu_{\rm c}$ (middle), and $\nu_m<\nu_a<\nu_c$ (bottom). Solid, dotted, dashed, and dash–dot lines represent $\alpha$ for coasting-phase outflows, BM decelerating relativistic jets, ST decelerating outflows, and decelerating outflows in DN phase, respectively. Different values of $p$ are color-coded: red, green, blue, and magenta for $p=2.01$, 2.5, 3, and 3.5. %When the relativistic outflow is in the deceleration phase, the CNM density profile cannot be derived from the CRs at high frequencies ($\nu_a < \nu_c < \nu < \nu_m$, $\nu_a < \nu_c < \nu_m < \nu$, $\nu_a < \nu_m < \nu_c < \nu$, and $\nu_m < \nu_a < \nu_c < \nu$). When the outflow is in the coasting phase, the CNM density profile cannot be derived from the CRs at low frequencies ($\nu < \nu_a < \nu_m < \nu_c$ and $\nu < \nu_m < \nu_a < \nu_c$).
}
\label{fig:CRs}
\end{figure*}

We plot $\alpha$ versus $k$ for various $p$ and different spectral regimes in Figure \ref{fig:CRs}. Solid, dotted, dashed, and dash–dot lines represent $\alpha$ for coasting-phase outflows, decelerating relativistic jets (BM), decelerating non-relativistic outflows (ST), and non-relativistic outflows in the deep-Newtonian phase (DN), respectively. Different values of $p$ are color-coded: red, green, blue, and magenta for $p=2.01$, 2.5, 3, and 3.5. The CRs for coasting-phase outflows with $\nu<\nu_a<\nu_m<\nu_c$ and $\nu<\nu_m<\nu_a<\nu_c$, and for decelerating relativistic outflows under $\nu_a<\nu_c<\nu<\nu_m$, $\nu_a<\nu_c<\nu_m<\nu$, $\nu_a<\nu_m<\nu_c<\nu$ and $\nu_m<\nu_a<\nu_c<\nu$, do not depend on $k$, meaning these CRs cannot constrain the CNM density profile. 

%Fortunately, most TDEs are in the coasting phase and are well observed at high frequencies. For a fixed dynamical phase, the CRs of individual spectral segments match for the \(\nu_a < \nu_m < \nu_c\) and \(\nu_m < \nu_a < \nu_c\) hierarchies, apart from the second segment. Since the characteristic radio frequencies of all known TDEs fall into these two hierarchies, our CR results are only weakly sensitive to spectral-case classification.

%---------------------------------------

\subsection{Method}

In this section, we describe the methods for obtaining the $k_{\rm CR}$ -index based on CR analysis, including spectral and light curve fitting procedures.

As the outflow evolves, the characteristic frequencies ($\nu_a,\nu_m, \nu_c$) vary, causing the radio spectral index \(\beta\) to transition between different spectral regimes over time. To derive the CNM density-profile index $k_{\rm CR}$ from the two indices $\alpha$ and $\beta$, we only use radio bands that exhibit negligible spectral evolution — and hence constant $\alpha$ and $\beta$ — within the observational time windows. These spectral indices are denoted $\alpha_{\rm CR}$ and $\beta_{\rm CR}$, respectively.

To do this, we first fit the radio spectra of each TDE at each observational epoch. A set of frequencies is sampled across the observed band, and light curves at these frequencies are built from the spectral-fitting outputs. Each light curve is fitted with a power law to yield the frequency-dependent temporal index $\alpha(\nu)$. We inspect the 
$\alpha(\nu) - \nu$ plane to locate the frequency range where $\alpha(\nu)$ flattens. This range marks a single spectral regime where the spectral index $\beta_{\rm CR}$ is defined. Averaging $\alpha(\nu)$ over this plateau gives $\alpha_{\rm CR}$. The CNM index $k$ is then evaluated via the expression for $k_{\rm CR}$ applicable in this regime.

\subsubsection{Spectral Fitting}
We fit the radio spectra of each TDE with the models developed by \citet{Granot2002}. Given that our observations target late-time radio emission and $\nu_{\rm c}$ is generally large, we consider only the regimes $\nu_{\rm a}<\nu_{\rm m}<\nu_{\rm c}$ and $\nu_{\rm m}<\nu_{\rm a}<\nu_{\rm c}$.

For $\nu_{\rm a}<\nu_{\rm m}<\nu_{\rm c}$, the spectrum can be described with 
\begin{equation}
    \begin{aligned}
        F_1 &= F_\nu(\nu_{\rm a}) \left[ \left(\frac{\nu}{\nu_{\rm a}}\right)^{s_1\beta_1}  + \left(\frac{\nu}{\nu_{\rm a}}\right)^{s_1\beta_2} \right]^{-1/s_1} \\
                          &\times \left[ 1 + \left(\frac{\nu}{\nu_{\rm m}}\right)^{s_2(\beta_3 - \beta_2)} \right]^{-1/s_2},
    \end{aligned}
\label{eq:specfit_amc}
\end{equation}
where $\beta_1 = -2$, $\beta_2 = -1/3$ and $\beta_3 = (p-1)/2$ are the spectra index, and $s_1=1.06$ and $s_2=1.76-0.38p$ are smoothing parameters. 
% (under the assumption that $k=2$)

For $\nu_{\rm m}<\nu_{\rm a}<\nu_{\rm c}$, the spectrum can be given by 
\begin{equation}
    \begin{aligned}
        F_2 &= F_{\nu}(\nu_{\rm m}) \left[ \left(\frac{\nu}{\nu_{\rm m}}\right)^2 \exp\left(-s_3\left(\frac{\nu}{\nu_{\rm m}}\right)^{2/3}\right) + \left(\frac{\nu}{\nu_{\rm m}}\right)^{5/2} \right] \\
                          &\times \left[ 1 + \left(\frac{\nu}{\nu_{\rm a}}\right)^{s_4(\beta_3 - \beta_2)} \right]^{-1/s_4},
    \end{aligned}
\label{eq:specfit_mac}
\end{equation}
where $\beta_2 = -5/2$, $\beta_3 = (p-1)/2$, and $s_3=3.63p-1.60$ and $s_4=1.25-0.18p$ are smoothing parameters. 
%(under the assumption that $k=2$)

For some TDEs, like AT2020vwl \citep{Goodwin2023-AT2020vwl, Goodwin2025-AT2020vwl}, ASASSN-14li \citep{Alexander16} and AT2020wjw \citep{Goodwin2024-J2344}, their radio emissions also include a host-galaxy component. We thus adopt
\begin{equation}
    F_{\nu,\rm host} = F_0\left( \frac{\nu}{1.4\ \rm GHz} \right)^{\alpha_0},
\label{eq:specfit_host}
\end{equation}
where $F_0$ is the flux density of the host galaxy at $\rm 1.4\ \,GHz$, $\alpha_0$ is the spectra index of the host galaxy. For these sources, the data are fitted with models of the form $F_1+F_{\nu,\rm host}$ (for $\nu_{\rm a}<\nu_{\rm m}$) or $F_2+F_{\nu,\rm host}$ (for $\nu_{\rm m}<\nu_{\rm a}$).

For each TDE, we fit the radio data spectra with two models, i.e., Equations (\ref{eq:specfit_amc}) and (\ref{eq:specfit_mac}), respectively. For each spectral model, we get the best fit using a Python implementation of Markov Chain Monte Carlo (MCMC), \texttt{emcee} \citep{emcee2013}. The parameter ranges are set as follows: $F_\nu \in [10^{-4}-100]$\,mJy, $F_0 \in [10^{-4}-2]$\,mJy, $\alpha_0 \in [-2-0]$, $p \in [2.0-4.0]$.
%, and the ranges of $\nu_{\rm a}$ and $\nu_{\rm m}$ are set according to the spectral properties of each individual source. 
For each MCMC run, we employed $\max(2n_{\rm dim}, 50)$ walkers ($n_{\rm dim}$ denotes the number of free parameters), 20000 steps, and discarded the first 10000 steps as burn-in. The radio SEDs and the MCMC fits for 26 TDEs are shown in Figures \ref{fig:sp_fit_1} and \ref{fig:sp_fit_2}. We select the preferred spectral model based on the BIC values derived from the best-fit results. The inferred values of $p$ as well as the preferred spectral model are listed in Table \ref{tab:calCR} (the notation a1/b1 corresponds to the regime $\nu_{\rm a}<\nu_{\rm m}<\nu_{\rm c}$, while a2/b2 denotes $\nu_{\rm m}<\nu_{\rm a}<\nu_{\rm c}$). For TDEs that include a host-galaxy component, the fitted $F_0$ and $\alpha_0$ are provided in Table \ref{tab:fitHost}.

%In our fitting procedure, the values of $F_0$, $\alpha_0$ and $p$ are constrained from a joint fit to all available spectra of the source. \textbf{We adopt the spectral model based BIC value from the best fit results.}. The resulting values for $p$ are listed in Table \ref{tab:calCR}, including the median value from the posterior distribution and the $1\sigma$ errors.

%16th and 84th percentiles, correspond to approximately $1\sigma$ errors.
%and the spectra index relevant to our CR analysis 

\subsubsection{Obtaining $\alpha_{\rm CR}$, $\beta_{\rm CR}$ and $k_{\rm CR}$}

Since the CRs are frequency dependent, we need to determine which spectral regime (and therefore $\beta$ value) and the $\alpha_{\rm CR}$ should be used for the $k_{\rm CR}$ calculation.

Based on the spectral-fitting results, we logarithmically sample 500 frequencies across the observational band for each SED and build the light curve for each frequency.
%computed the corresponding model fluxes. %Repeating this procedure for all SEDs of the TDE yields a light curve at every sampled frequency. By fitting each light curve with a power-law function, we can obtain the frequency-dependent distribution of temporal indices. 
Subsequently, we perform a power-law fit on each light curve to obtain the frequency-dependent distribution of the temporal index $\alpha(\nu)$. The $\alpha(\nu) - \nu$ planes for 26 TDEs are presented in Figures \ref{fig:alpha_cr_1} and \ref{fig:alpha_cr_2}.
%the light curves at 500 sampled frequencies to obtain the frequency-dependent distribution of the temporal index.

Inspecting Figures \ref{fig:alpha_cr_1} and \ref{fig:alpha_cr_2},
%the $\alpha(\nu) - \nu$ planes for the sampled TDEs, 
we find that $\alpha(\nu)$ flattens at lower frequency ($\nu< \min(\nu_{\rm a},\nu_{\rm m})$ and therefore $\beta_{\rm CR}=-2$) or at higher frequency ($\nu> \max(\nu_{\rm a},\nu_{\rm m})$ and therefore $\beta_{\rm CR}=(p-1)/2$). The former case generally arises in the relativistic jet, while the latter typically occurs for non-relativistic outflows. 
%To show how we obtain $\alpha_{\rm CR}$, we present the radio SED fits (top) as well as the $\alpha(\nu) - \nu$ curves (bottom) for Swift J1644+57 (the well-known jetted TDE) and CNSS J0019+00 (a TDE with non-relativistic outflow) in Figure \ref{fig:method}. This range marks a single spectral regime where the spectral index $\beta_{\rm CR}$ is defined. Averaging $\alpha(\nu)$ over this plateau gives $\alpha_{\rm CR}$. The CNM index $k$ is then evaluated via the expression for $k_{\rm CR}$ applicable in this regime.}

\begin{figure*}[htp]
\center
\includegraphics[width=1.0\textwidth]{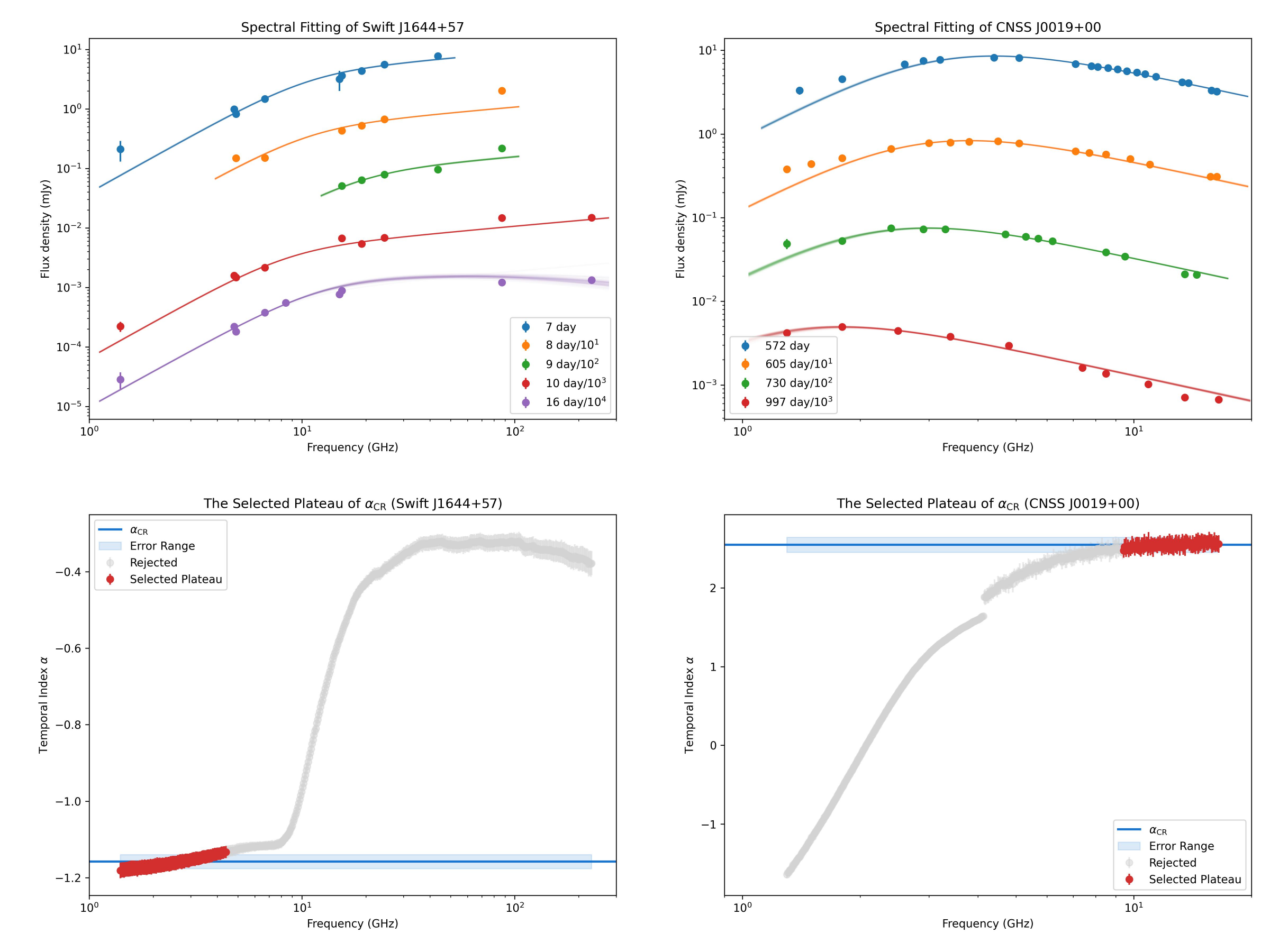}
\caption{Top: Radio SED fitting results for Swift J1644+57 and CNSS J0019+00. The lines represent the fits from our MCMC modeling. Bottom: The $\alpha(\nu) - \nu$ plane obtained with the power-law fit to the light curves of each frequency built from the spectral-fitting outputs. $\alpha(\nu)$ flattens at lower frequency ($\nu< \nu_{\rm a}$ and then $\beta_{\rm CR}=-2$) for Swift J1644+57, and at higher frequency ($\nu> \nu_{\rm a}$ and then $\beta_{\rm CR}=(p-1)/2$) for CNSS J0019+00. Averaging $\alpha(\nu)$ over this plateau (red) gives $\alpha_{\rm CR}$ (the blue lines). The blue shaded regions show the error range.}
\label{fig:method}
\end{figure*}

To show how we obtain $\alpha_{\rm CR}$, $\beta_{\rm CR}$ and $k_{\rm CR}$, we take Swift J1644+57 (the well-known jetted TDE) and CNSS J0019+00 (a TDE with non-relativistic outflow) for example, see Figure \ref{fig:method}. %Their radio TDEs are presented in the top panel of  Figure \ref{fig:method}. 
The spectral fitting with MCMC for both sources prefers the regime $\nu_{\rm a}<\nu_{\rm m}<\nu_{\rm c}$, The $p$-values are $2.28\pm0.17$ for Swift J1644+57 and $3.04\pm0.02$ for CNSS J0019+00. $\alpha(\nu)$ flattens at $\nu<\nu_{\rm a}$ ($\nu>\nu_{\rm m}$) for Swift J1644+57 (CNSS J0019+00), and then we have $\beta_{\rm CR}=-2$ ($\beta_{\rm CR}=(p-1)/2=1.02\pm0.01$). We initialize the plateau with the three lowest frequency points ($\nu< \nu_{\rm a}$) for Swift J1644+57, and highest frequency points ($\nu> \nu_{\rm m}$) for CNSS J0019+00. The median value and median absolute deviation are calculated for the current plateau. We then extend the plateau to higher (lower) frequencies for Swift J1644+57 (CNSS J0019+00) until the deviation from the median-$\alpha$ substantially exceeds the median absolute deviation. Finally, we take the weighted average of $\alpha$ across the new plateaus (see the red lines in Figure \ref{fig:method}) as $\alpha_{\rm CR}=-1.16$ for Swift J1644+57 and $\alpha_{\rm CR}=2.55$ for CNSS J0019+00 (see the blue lines in Figure \ref{fig:method}). The uncertainty is given by the standard error of the weighted mean. In terms of dynamics, relativistic TDEs tend to reside in the BM phase \citep{Wang2014}, whereas thermal TDEs are more likely to be in the coasting phase \citep{Matsumoto+Piran2024,Lei2026CNM,Bu2026}. Therefore, we adopt the expression $k_{\rm CR} =\frac{4\alpha-6\beta}{\alpha-3\beta}$ ($\nu<\nu_{\rm a}$ in the BM deceleration phase, see Table \ref{tab:CR_BM}) for Swift J1644+57, and $k_{\rm CR} =\frac{2\alpha+6}{3+\beta}$ ($\nu_{\rm a}<\nu<\nu_{\rm m}$ in the coasting phase, see Table \ref{tab:CR_coasting}) for CNSS J0019+00. Substituting $\beta_{\rm CR} =-2$ ($\beta_{\rm CR}=(p-1)/2=1.02\pm0.01$) and $\alpha_{\rm CR}=-1.16\pm0.02$ ($\alpha_{\rm CR}=2.55\pm0.09$) into the $k_{\rm CR}$ expressions, we have the estimation of the CNM density profile index $k_{\rm CR}=2.27\pm0.03$ ($k_{\rm CR}=2.76\pm0.05$) for Swift J1644+57 (CNSS J0019+00).

%We prefer to obtain $\alpha_{\rm CR}$ by fitting the light curves of the lower ($< \min(\nu_{\rm a},\nu_{\rm m})$) or higher ($> \max(\nu_{\rm a},\nu_{\rm m})$) frequencies, since these frequencies are usually located at the same spectral regime (we do not need to consider the transition of $\alpha$ due to crossing of the peak frequency, i.e., $\nu_{\rm a}$ or $\nu_{\rm m}$). For each frequency, we obtain a temporal index $\alpha$ from its light curve. The resulting $\alpha_{\rm CR}$ of a TDE is the average of all these temporal indices.
%Our synchrotron model indicates that the dependence of temporal indices on frequency contains several plateau‑like regions. We selected the segment of the curve that exhibits the most stable plateau and computed the average of all temporal indices within this region to obtain $\alpha_{\rm CR}$. We then identified the spectral phase and its associated spectral index $\beta_{\rm CR}$ from the frequency region corresponding to the plateau. 

%Substituting $\alpha_{\rm CR}$ and $\beta_{\rm CR}$ into the analytical expression of CR allows us to infer the density profile of the CNM.

%Given that the observations made in the bands of existing radio equipment fall mostly within the high-frequency of the TDE radio spectrum, and the decay of $\nu_a$ and $\nu_m$, it is challenging to obtain a stable plateau in the low-frequency part without frequent observations in the early time. Therefore, for most TDEs, the most stable plateau are located in high frequency regime. % and we search for the plateau forward from the higher to lower frequencies.

\section{Results and Discussions}
\label{sec:results+discuss}

The radio SEDs (as well as the MCMC fits) and the $\alpha(\nu) - \nu$ planes for the sample of 26 TDEs are shown in Figures \ref{fig:sp_fit_1} and \ref{fig:sp_fit_2}. The inferred values of $p$ (as well as the preferred spectral model, the notation a1/b1 corresponds to the regime $\nu_{\rm a}<\nu_{\rm m}<\nu_{\rm c}$, while a2/b2 denotes $\nu_{\rm m}<\nu_{\rm a}<\nu_{\rm c}$), $\alpha_{\rm CR}$, $\beta_{\rm CR}$ and $k_{\rm CR}$ are listed in Table \ref{tab:calCR}. We compute $k_{\rm CR}$ in the BM phase for relativistic jetted TDEs and in the coasting phase for TDEs with non-relativistic outflows. For comparison, we also present the $k_{\rm eq}$-value obtained from equipartition method. For TDEs that include a host-galaxy component, the fitted $F_0$ and $\alpha_0$ are provided in Table \ref{tab:fitHost}.

%Applying these criteria, we yield a sample of 26 TDEs to constrain the CNM density profile index with the CR analysis. 
There are several typical cases: (1) Sw J1644+57 and AT2022cmc are the two TDEs in our sample for which the CR analysis requires a relativistic deceleration phase. (2) CNSS J0019+00 represents the typical behavior of most radio TDEs, with a non‑relativistic outflow in the coasting phase. (3) Five TDEs, i.e., AT2020vwl, ASASSN-14li, AT2020wjw, AT2019ahk and AT2019azh, exhibit radio emissions that include both the synchrotron component and a contribution from its host galaxy. 
%The spectra fitting of these three TDEs are presented in Figure \ref{fig:sp}. 

For AT2022cmc, most of its SEDs are in the low‑frequency regime (corresponding to a spectral index of $\beta_{\rm CR}=-2$), we obtain the mean temporal index $\alpha_{\rm CR}=-0.72 \pm 0.02$ from the light curves of this low-frequency range ($\nu<\nu_a<\nu_m<\nu_c$). Assuming the outflow is in the BM deceleration phase, the CR analysis yields a CNM density profile of $k_{\rm CR}=1.23 \pm 0.06$, which is consistent with our previous work \citep{Zhou2024} and also the value obtained with equipartition method. 

Recently, \citet{An2026} studied the delayed radio flare from a supermassive black hole binary candidate SDSS J143016.05+230344.4. The equipartition analysis indicated a steep CNM density profile $k=1.7$. Therefore, such sources can also provide direct constraints on sub-parsec CNM density profile.

\begin{table*}[htbp]
\centering
\begin{threeparttable} 
\caption{Parameters of the CR analysis.}
\label{tab:calCR}

\begin{tabular}{cccccc}
\toprule
TDE&$\beta_{CR}$ & $p$ &$\alpha_{CR}$  & $k_{\rm CR}$&$k_{\rm eq}$\\ %& $k_{\rm PyFRS}$ \\
\midrule
    Swift J1644+57&$-2$&$2.28\pm0.17$&$-1.16\pm0.02$&$2.27\pm0.03$\tnote{a1}&$1.5$ \citep{Eftekhari18}\\%2nd
\cmidrule(lr){1-6}
    AT2022cmc &$-2$&$2.15\pm0.14$&$-0.72\pm0.02$&$1.23\pm0.06$\tnote{a1}&$1.5-2$ \citep{Matsumoto2023}\\%1st
\cmidrule(lr){1-6}
    ASASSN-15oi&$1.15\pm0.13$&$3.31\pm0.26$&$3.70\pm0.10$&$3.22\pm0.11$\tnote{b1}&$-$\tnote{c}\,\, \citep{Hajela2025-15oi}\\%2nd
\cmidrule(lr){1-6}
    AT2024tvd &$1.43\pm0.14$&$3.85\pm0.28$&$4.37\pm0.09$&$3.33\pm0.11$\tnote{b1} %, $4.58\pm0.06$\tnote{b}
    &$3.20$\tnote{d}\,\, \citep{Sfaradi2025-AT2024tvd}\\%3(delayed)
\cmidrule(lr){1-6}
    AT2020vwl\tnote{*}&$1.47\pm0.04$&$3.95\pm0.08$&$-0.61\pm0.33$&$1.07\pm0.15$\tnote{b1}&$1.90$\tnote{d}\,\, \citep{Goodwin2025-AT2020vwl}\\%2nd
\cmidrule(lr){1-6}
    ASASSN-14li\tnote{*}&$1.30\pm0.05$&$3.60\pm0.10$&$1.68\pm0.12$&$2.18\pm0.06$\tnote{b1}&2.5 \citep{Alexander16}\\
\cmidrule(lr){1-6}
    AT2020wjw\tnote{*} &$1.47\pm0.04$&$3.94\pm0.09$&$-0.04\pm0.06$&$1.32\pm0.03$\tnote{b2}& $1.4$ \citep{Goodwin2024-J2344}\\
\cmidrule(lr){1-6}
    ASASSN-14ae&$0.51\pm0.02$&$2.03\pm0.03$&$-5.13\pm0.07$&$-1.21\pm0.04$\tnote{b1}&$-$\tnote{c}\,\, \citep{Cendes2024}\\
\cmidrule(lr){1-6}
    AT2018hyz&$0.50\pm0.01$&$2.01\pm0.01$&$-4.20\pm0.02$&$-0.69\pm0.01$\tnote{b2}&$0.5$ \citep{Cendes2026_AT2018hyz_radio}\\
\cmidrule(lr){1-6}
    AT2019ahk\tnote{*}&$1.26\pm0.14$&$3.52\pm0.29$&$-0.34\pm0.06$&$1.25\pm0.05$\tnote{b2}&$1.48$\tnote{d}\,\, \citep{Christy2024ASASSN-19bt}\\
\cmidrule(lr){1-6}
    AT2019avd&$1.07\pm0.26$&$3.14\pm0.52$&$0.00\pm0.01$&$1.48\pm0.09$\tnote{b2}&$-$\tnote{c}\\
\cmidrule(lr){1-6}
    AT2019azh\tnote{*}&$1.50\pm0.01$&$3.99\pm0.01$&$0.23\pm0.12$&$1.43\pm0.05$\tnote{b1}&$0.24$\citep{Goodwin2022-19azh}\\
\cmidrule(lr){1-6}
    AT2019dsg&$0.86\pm0.06$&$2.72\pm0.13$&$0.01\pm0.02$&$1.56\pm0.03$\tnote{b2}&$1.7$ \citep{Cendes2021-19dsg}\\
\cmidrule(lr){1-6}
    AT2019ehz&$1.46\pm0.06$&$3.92\pm0.12$&$3.81\pm0.09$&$3.05\pm0.06$\tnote{b2}&$19.50$\tnote{d}\,\, \citep{Cendes2024}\\
\cmidrule(lr){1-6}
    AT2019eve&$1.42\pm0.05$&$3.85\pm0.10$&$0.26\pm0.10$&$1.47\pm0.05$\tnote{b1}&$6.16$\tnote{d}\,\, \citep{Cendes2024}\\
\cmidrule(lr){1-6}
    AT2019teq&$0.85\pm0.10$&$2.71\pm0.20$&$0.18\pm0.16$&$1.65\pm0.09$\tnote{b2}&$-$\tnote{c}\,\, \citep{Cendes2024}\\
\cmidrule(lr){1-6}
    AT2020opy&$1.20\pm0.19$&$3.41\pm0.38$&$0.01\pm0.04$&$1.43\pm0.07$\tnote{b2}&$1.54$\tnote{d}\,\, \citep{Goodwin2023-AT2020opy}\\
\cmidrule(lr){1-6}
    AT2020vdq&$1.05\pm0.05$&$3.10\pm0.10$&$0.03\pm0.08$&$1.50\pm0.04$\tnote{b2}&$-$\\
% \cmidrule(lr){1-6}
%     AT2022dsb&$0.80\pm0.13$&$2.60\pm0.27$ (2.52)&$0.66\pm0.02$&$1.93\pm0.07$\tnote{a}&$-$\\
\cmidrule(lr){1-6}
    CNSS J0019+00&$1.02\pm0.01$&$3.04\pm0.02$&$2.55\pm0.09$&$2.76\pm0.05$\tnote{b2}&$2.5$ \citep{Anderson2020-CNSS0019}\\
\cmidrule(lr){1-6}
    eRASSt J011431-593654&$1.30\pm0.28$&$3.59\pm0.55$&$0.09\pm0.12$&$1.44\pm0.11$\tnote{b1}&$-$\tnote{c}\,\, \citep{Goodwin2025-22eROSAT_sample}\\
\cmidrule(lr){1-6}
    eRASSt J210858-562832&$1.15\pm0.25$&$3.31\pm0.49$&$0.08\pm0.11$&$1.48\pm0.10$\tnote{b2}&$-$\tnote{c}\,\, \citep{Goodwin2025-22eROSAT_sample}\\
\cmidrule(lr){1-6}
    IGR J12580+0134&$1.49\pm0.01$&$3.99\pm0.01$&$0.01\pm0.05$&$1.34\pm0.02$\tnote{b2}&$-$\tnote{c}\\
\cmidrule(lr){1-6}
    iPTF 16fnl&$1.13\pm0.34$&$3.26\pm0.67$&$0.09\pm0.12$&$1.49\pm0.14$\tnote{b2}&$-$\tnote{c}\,\, \citep{Cendes2024}\\
\cmidrule(lr){1-6}
    PS16dtm&$1.46\pm0.06$&$3.91\pm0.13$&$0.21\pm0.17$&$1.44\pm0.08$\tnote{b1}&$-$\tnote{c}\,\, \citep{Cendes2024}\\
\cmidrule(lr){1-6}
    WTP14adeqka&$1.18\pm0.01$&$3.36\pm0.01$ &$-1.98\pm0.02$&$0.49\pm0.01$\tnote{b1}&$3.33$\tnote{d}\,\, \citep{Golay2025-WTP14adeqka}\\
\cmidrule(lr){1-6}
    XMMSL1 J0740-85&$1.47\pm0.04$&$3.95\pm0.08$&$0.31\pm0.24$&$1.48\pm0.11$\tnote{b1}&$2.35$ \citep{Alexander2017-XMMSL1}\\
\bottomrule
\end{tabular}
\begin{tablenotes}
\item[a] The CNM density profile at BM deceleration phase, where a1 for $\nu_{\rm a}<\nu_{\rm m}<\nu_{\rm c}$ and a2 for $\nu_{\rm m}<\nu_{\rm a}<\nu_{\rm c}$. 
\item[b] The CNM density profile at Coasting phase, where b1 for $\nu_{\rm a}<\nu_{\rm m}<\nu_{\rm c}$ and b2 for $\nu_{\rm m}<\nu_{\rm a}<\nu_{\rm c}$.
% \item[c] The CNM density profile at Newtonian phase.
\item[c] Due to the lack of peak SED observations, it is difficult to constrain $k_{\rm eq}$.
\item[d] $\rm k_{\rm eq}$ is obtained by performing linear fitting on the equipartition results from other work.
\item[*] TDE radio SEDs with a host-galaxy component.
\end{tablenotes}
\end{threeparttable}
\end{table*}

%%%%%%%%%%%%%%%%%%%%%%%%%%%%%%%%%%%%%%%%%%%%%%%%%%%%%%%%%%%%%%%%%%%

%%%%%%%%%%%%%%%%%%%%%%%%%%%%%%%%%%%%%%%%%%%%%%%%%%%%%%%%%%%
\begin{table*}[htbp]
\centering
\begin{threeparttable} 
\caption{\textbf{Host-galaxy component}.}
\label{tab:fitHost}

\begin{tabular}{cccccc}
\toprule
TDE&AT2020vwl & ASASSN-14li& AT2020wjw & AT2019ahk & AT2019azh  \\ %& $k_{\rm PyFRS}$ \\

\cmidrule(lr){1-6}    $F_0$(mJy)&$0.63\pm0.11$&$1.25\pm0.15$&$0.55\pm0.24$&$1.25\pm0.26$&$1.38\pm0.02$\\%2nd
\cmidrule(lr){1-6}
    $\alpha_0$ &$-1.16\pm0.12$&$-1.23\pm0.10$&$-1.22\pm0.28$&$-0.92\pm0.28$&$-0.85\pm0.01$ \\

\bottomrule
\end{tabular}

\end{threeparttable}
\end{table*}

%%%%%%%%%%%%%%%%%%%%%%%%%%%%%%%%%%%%%%%%%%%%%%%%%%%%%%%%%%%%%%%%%%%

\begin{figure*}[htp]
\center
\includegraphics[width=1.0\textwidth]{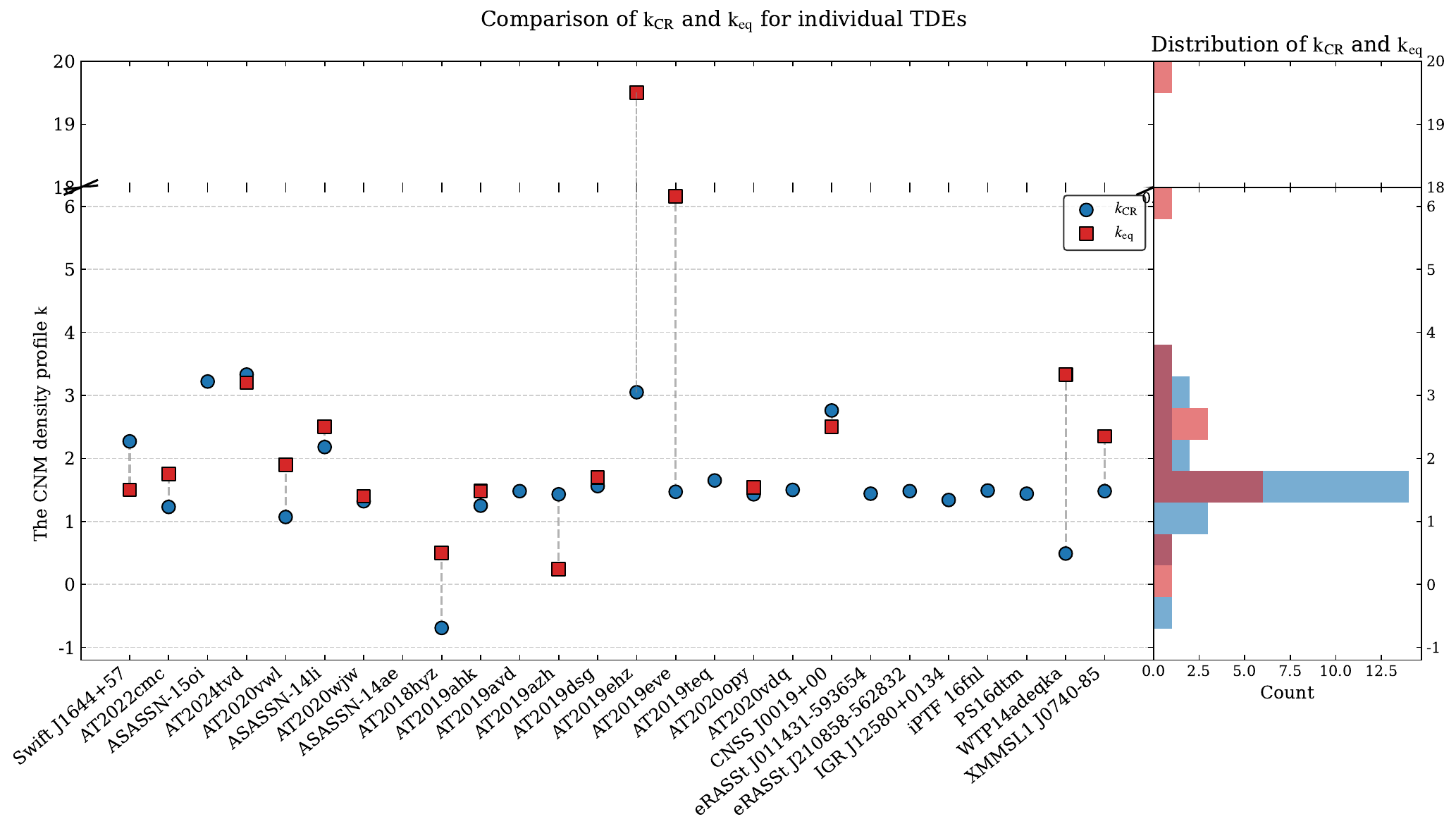}
\caption{Left panel: The CNM density profile for our sample of 26 TDEs, as inferred from CRs and equipartition methods. Right panel: The distributions of the CNM density profile $k_{\rm CR}$ and $k_{\rm eq}$, together with the best-fit Gaussian curves.}
\label{fig:distribution}
\end{figure*}

\subsection{Negative Profile Index}
Two TDEs in our sample, i.e., ASASSN-14ae and AT2018hyz exhibit negative CNM density slopes ($k<0$), indicating an unreasonable inverted density profile. %The results derived from CRs method of these TDEs are summarized below.

ASASSN-14ae has two SEDs at 2696 and 3243\,day, suggesting an electron spectral index of $p= 2.03 \pm 0.03$, The average temporal index is $\alpha_{\rm CR}= -5.13 \pm 0.07$, and the corresponding spectral index is $\beta_{\rm CR}= 0.51 \pm 0.02$. We obtain a CNM density profile of $k_{\rm CR}= -1.21 \pm 0.04$ during the coasting phase by CR analysis. AT2018hyz was modeled using 16 SEDs spanning 1125–2160\,days, and the spectral fitting yields $p= 2.01 \pm 0.01$. The average temporal index is $\alpha_{\rm CR}= -4.20 \pm 0.02$, and the corresponding spectral index is $\beta_{\rm CR}=(p-1)/2=0.50 \pm 0.01 $. We obtain a CNM density profile of $k_{\rm CR}= -0.69 \pm 0.01$ during the coasting phase by CR analysis. 
%eRASSt J011431-593654 was modeled by fitting two SEDs at 1224 and 1415\,days, yielding $p=3.42 \pm 0.60 $, The light curves at high-frequencies show an average temporal index of $\alpha_{\rm CR}= -3.77 \pm 0.29$, with a corresponding spectral index of $\beta_{\rm CR}=(p-1)/2= 1.21 \pm 0.30$. By substituting these two parameters into the CR expression, we obtain a CNM density profile of $k_{\rm CR}= -0.36 \pm 0.14$ during the coasting phase.

\citet{Matsumoto+Piran2024} explored a model of a changed CNM density profile that initially decreases as a power-law with radius but tends to a constant value beyond the Bondi radius, which can explain the delayed radio emission. In this model, after entering a constant-density medium, the growth of radio emission over time with $L_\nu\propto t^3$, which is determined by the cubic relation between the number of swept-up electrons and the radius ($N_{\rm e}\propto R^3$) and the fact that the outflow propagates at a constant velocity before deceleration ($R\propto t$). Therefore, this model struggles to explain the steep rise ($\alpha_{\rm CR} < -3$) light curve. 
%as in AT2018hyz. 
%Therefore, the steeply rising light curves ($\alpha_{CR} < -3$) of the three TDEs with negative CNM density profiles imply the existence of other mechanisms.

Recently, several works suggested that the late-time fast rising can be produced when the outflow encounters surrounding gaseous clouds or a torus \citep{Mou2021,Mou2022,Bu2023,Bu2026, Lei2024-torus,Lei2026CNM, Zhuang2025}. 
%These negative density profile index indicates that the ambient density of these sources deviates locally from a simple powerlaw decay and instead begins to increase, like the outflow encounters surrounding gaseous clouds or a torus \citep{Mou2021,Mou2022,Bu2023,Lei2024-torus,Zhuang2025}. 
\citet{Lei2024-torus} proposed a model investigating the interaction between the unbound debris and the external dusty torus commonly found in galactic nuclear regions, particularly in active galactic nuclei. Using three-dimensional hydrodynamic simulations with the \texttt{Athena++} code, their results show that the debris-torus collision generates strongly shocked material, and the resulting radio emission exhibits a characteristic light curve with a sharp rise and slow decay. \citet{Bu2026} further investigated the effects of the angular distribution of outflow properties on the radio light curve. \citet{Zhuang2025} proposed an outflow-cloud interaction model to explain the late-time radio flare in AT2018hyz. In his model, a sub-relativistic outflow generated at early stages collides with discrete, dense clouds located on scales of 0.1–1 pc during its propagation, forming bow shocks that produce radio flares through synchrotron radiation. This naturally explains both the time delay (several years) of the late flare in AT2018hyz and the steep rise of its light curve. Therefore, the negative density profile indices might be the encounter of dense medium such as a cloud or torus.

\citet{Matsumoto2023MNRAS, Sato2024,Cendes2026_AT2018hyz_radio} showed that the late-time radio rebrightening (like ASASSN-14ae, ASASSN-15oi, AT2018hyz and AT2019dsg) can be explained by the off-axis two-component jet model, in which the late-time rebrightening arises from the off-axis view of a decelerating narrower jet. Very long baseline interferometry (VLBI) imaging can provide the most direct way to distinguish between these scenarios \cite{Matsumoto2026ApJ}.

\citet{Cendes2022-AT2018hyz} suggested that the steep rise in AT2018hyz can not be explained in the scenario of an outflow launched at the time of the disruption, and should instead be a delayed outflow launch. The late time SED properties may involve an internal shocks within the outflow. For AT202vwl and ASASSN-15oi, the radio re-flaring episode is thought to correspond to a delayed outflow switching on \citep{Goodwin2026,Sato2025}. Such delayed outflow is expected if a precessing jet propagates in the envelope/wind of TDEs \citep{Teboul2023,Yuan2026}.

%Therefore, both unbound debris-torus interaction and outflow–cloud interaction can produce delayed, rapidly rising radio flares, and the obtained negative density profile is self-consistent with the steep rise observed in the source itself.

\subsection{Comparing with Equipartition Method}
In table \ref{tab:calCR}, we also present the equipartition values $k_{\rm eq}$ obtained from the literature. The comparison of these two values, i.e., $k_{\rm CR}$ and $k_{\rm eq}$, and the distributions are shown in Figure \ref{fig:distribution}.

%, including the values directly given in previous work and the values we obtained by fitting the results from previous work.
%By comparing the CNM density profiles inferred from the CR method with those derived from the equipartition method, we present the CNM density profile for each TDE obtained from both approaches, along with their distributions, in Figure \ref{fig:distribution}. Overall, 

As we can see, the results of the two methods are generally consistent, with the CNM density profiles predominantly concentrated around $k=2$. However, two TDEs, namely AT2019ehz and AT2019eve, exhibit significant discrepancies between the two methods.

For CR analysis, AT2019ehz uses $2$ SEDs on $970$ and $1262$\,days. We obtain $p= 3.92 \pm 0.12$ by the spectral fitting. The average temporal index is $\alpha_{\rm CR}= 3.81 \pm 0.09$, with a corresponding spectral index of $\beta_{\rm CR}=(p-1)/2= 1.46 \pm 0.06$. By substituting these two parameters into the CR expression, we obtain a CNM density profile of $k_{\rm CR}=3.05 \pm 0.06$ during the coasting phase. For AT2019eve, we use $3$ SEDs that include $945$, $1092$ and $1325$\,days. We then get $p=3.85 \pm 0.10 $, an average temporal index of $\alpha_{\rm CR}= 0.26 \pm 0.10$, and the corresponding spectral index of $\beta_{\rm CR}=(p-1)/2= 1.42 \pm 0.05$. With CR expression, we obtain a CNM density profile of $k_{\rm CR}= 1.47 \pm 0.05$ during the coasting phase.

%The results of these TDEs are summarized here. 
%AT2019ehz has $2$ SEDs including $970$ and $1262$\,days, these SEDs are fitted by $p= 3.94 \pm 0.06$. The plateau phase shows an average temporal index of $\alpha_{CR}= 3.72 \pm 0.09$, corresponding to a spectral index of $\beta_{CR}=(p-1)/2= 1.47 \pm 0.03$. By substituting these two parameters into the CR expression, we obtain a CNM density profile of $k_{CR}=3.00 \pm 0.05$ during the coasting phase. AT2019eve has $3$ SEDs including $945$, $1092$ and $1325$\,days, these SEDs are fitted by $p=3.04 \pm 0.06 $. The plateau phase shows an average temporal index of $\alpha_{CR}= 0.75 \pm 0.09$, corresponding to a spectral index of $\beta_{CR}=(p-1)/2= 1.02 \pm 0.03$. By substituting these two parameters into the CR expression, we obtain a CNM density profile of $k_{CR}= 1.87 \pm 0.04$ during the coasting phase.

%In equipartition analysisi, AT2019ehz and AT2019eve exhibit notable discrepancies, 

The CNM density slopes derived from the equipartition method are $k_{\rm eq}=19.50$ and $k_{\rm eq}=6.16$ for AT2019ehz and AT2019eve, respectively \citep{Cendes2024}. The large discrepancies with the results of CR analysis lie in the fact that poor data are used. For both cases, the values of $k_{\rm eq}$ are obtained from only two or three time epochs. There are large uncertainties induced, since $k_{\rm eq}$ is usually variable with radius (and time). The equipartition method highly relies on the measurements of the peak frequency ($\nu_{\rm a}$ or $\nu_{\rm m}$), which introduce additional error with the poor SEDs in AT2019ehz and AT2019eve. 

%These values further highlight the substantial differences between the two methods in these cases. This may be attributed to the fact that, in equipartition analysis, the SED fits are performed independently for each epoch, without coupling between parameters. As a result, the fitting outcomes at different epochs are entirely independent of one another. For instance, in the case of AT2019ehz, calculating the equivalent velocity from the equipartition radius and the times corresponding to its two equipartition-derived SEDs yields $\beta \approx 0.0025$, which is substantially lower than the equipartition-inferred values of $\beta = 0.017$ and $\beta = 0.014$. Moreover, for both AT2019ehz and AT2019eve, the SED peaks at certain epochs are not distinctly pronounced, potentially introducing inaccuracies in peak fitting and leading to a derived CNM density profile that deviates from the expectations based on CRs analysis.

\section{Summary}
\label{sec:summary}
With the rapid increase in radio observations of TDEs, and given that their radio emission is generally produced by the interaction between an outflow and the surrounding circumnuclear medium (CNM), radio‑detected TDEs have become a powerful probe of the CNM density profile. In this work, we collect the TDEs with radio observations and constrain the CNM density profile with CR analysis. The main results of this work are summarized as follows:
\begin{enumerate}
    \item We obtain a sample of 53 TDEs with long‑term radio observations, including 14 X-ray (or gamma-ray) selected TDEs, 34 optical selected TDEs, 3 infrard selected TDEs and 2 radio selected TDEs. There are only five TDEs with a relativistic jet. For AT2019teq and AT2020vdq, we reduce the VLA data following standard procedures using the CASA package. 
    This TDE sample enables us to carry out a comprehensive study of the CNM of dormant SMBHs.
    %representing the largest collection of radio‑emitting TDEs available to date and enabling a comprehensive statistical investigation.
    \item We present the CR expressions for the synchrotron emission from the interaction of outflow with an arbitrary CNM, establishing the relationship between the CNM density profile index $k$ and the temporal index $\alpha$ (given the spectral index $\beta$). This provides a theoretical framework for constraining CNM properties using radio data.
    \item We select 26 TDEs from the sample that suitable for CR analysis, and obtain CNM profile with the CR method. We find that our results are generally consistent with that from equiparpition method, indicating that CR analysis remains a powerful diagnostic of the CNM profile.
    %Most TDEs contain  to these sources to infer their CNM density distributions.
\end{enumerate}
This study presents an efficient method for estimating the CNM density profile.

%the most extensive radio TDE sample currently available and offers a practical and efficient method for estimating the CNM density profile. 
Future radio facilities, like FAST Core Array \citep{FASTCoreA2024} and SAK \citep{Shu2026SKAbook,An2026SKAbook}, with higher sensitivity and resolution are expected to collect rich radio data for TDEs, which will advance our understanding of the CNM environments of quiescent SMBHs.
%deliver larger, more distant, and more diverse samples of radio TDEs, to advance our understanding of the CNM environments of quiescent supermassive black holes.

\section*{acknowledgments}
We thank the participants of the TDE FORUM (Full-process Orbital to Radiative Unified Modeling) online seminar series for their inspiring discussions. We thank Patrik Mil\'an Veres for sharing the the multi-band ATCA fluxes and the EVN observation's result. We are very grateful to Yanan Wang, Rongfeng Shen, Xinwen Shu, Bing Zhang, Erlin Qiao, Tao An, Agnieszka Janiuk, Chichuan Jin and Fangkun Peng for their helpful discussions. This work is supported by the National Natural Science Foundation of China under grants 12473012 and 12533005, the National Key R\&D Program of China (Nos. 2023YFC2205901), and the Fundamental Research Funds for the Central Universities, HUST (No. YCJJ20252115). The authors acknowledge Beijing PARATERA Tech CO., Ltd. for providing HPC resources that have contributed to the research results reported within this paper.

\bibliography{ref}{}
\bibliographystyle{aasjournal}

\begin{appendix}
\renewcommand{\thetable}{A\arabic{table}}
\setcounter{table}{0}

\renewcommand{\thefigure}{B\arabic{figure}}
\setcounter{figure}{0}

\section{TDE with Detected Radio Emission} \label{app:data}

%%%%%%%%%%%%%%%%%%%%%%%%%%%%%%%%%%%%%%%%%%%%%%%%%%
\begin{ThreePartTable} 
\begin{TableNotes} 
\item[a] The initial date of the optical rebrightening \citep{Wang2025-2020afhd}. 
\item[b] The time of the peak eROSITA X-ray flux measurement.
\item[c] The time of optical peak.
% \item[e] The radio data in \citet{Somalwar2025-AT2020vdq} have not been made publicly available.
\item[d] We thank Patrik Mil\'an Veres for sharing the the multi-band ATCA fluxes and the EVN observation's result. %The radio data have not been released.
\end{TableNotes}

\begin{longtable}{ccccc}
\caption{The informations of TDEs with radio detections.} \label{tab:all_radio_tde}\\
\toprule
% \multicolumn{5}{c}{The basic features} \\
Object & Redshift & Discovery band & Discovery time & Data reference \\
\midrule
\endfirsthead

\caption[]{The radio TDEs samples (continued)}\\
\toprule
\multicolumn{5}{c}{The basic features (continued)} \\
Object & Redshift & Discovery band & First detected time & Radio data reference \\
\midrule
\endhead
% \bottomrule
% \insertTableNotes
\endfoot
\endlastfoot
%    \multirow{1}{*}{Swift J1644+57 }& \multirow{1}{*}{0.354}& \multirow{1}{*}{$\gamma$ ray}& \multirow{1}{*}{2011 March 25}& \citet{Berger2012,Zauderer2013,Eftekhari18,Cendes2021-1644} \\     
   \multirow{4}{*}{Swift J1644+57 }& \multirow{4}{*}{0.354}& \multirow{4}{*}{$\gamma$ ray}& \multirow{4}{*}{2011 March 25}& \citet{Berger2012} \\ & & & &\citet{Zauderer2013}\\ & & & & \citet{Eftekhari18}\\& & & & \citet{Cendes2021-1644}\\
\cmidrule(lr){1-5}
    \multirow{1}{*}{Swift J1112-82}& \multirow{1}{*}{0.8901}& \multirow{1}{*}{$\gamma$ ray}& \multirow{1}{*}{2011.06.16-19}& \citet{Brown2017-SwiftJ1112-82} \\
\cmidrule(lr){1-5}
    \multirow{3}{*}{Swift J2058+05}& \multirow{3}{*}{1.1853}& \multirow{3}{*}{$\gamma$ ray}& \multirow{3}{*}{2011.05.17-20}&\citet{Cenko2012-SwJ2058+05} \\ & & & & \citet{Pasham2015-SwJ2058+05} \\& & & & \citet{Brown2017-SwiftJ1112-82} \\
\cmidrule(lr){1-5}
    \multirow{4}{*}{AT2022cmc}& \multirow{4}{*}{1.19325}& \multirow{4}{*}{Optical}& \multirow{4}{*}{2022 February 11}& \citet{Andreoni2022}\\ & & & & \citet{Pasham2023} \\& & & &\citet{Rhodes2023}\\&&&&\citet{Rhodes2025}\\
\cmidrule(lr){1-5}
    \multirow{2}{*}{EP250702a}& \multirow{2}{*}{1.036}& \multirow{2}{*}{X-ray}& \multirow{2}{*}{2025 July 2}& \citet{Li2025-EP250702a}\\
    &&&&\citet{Levan2025-0702a}\\
\cmidrule(lr){1-5}
    \multirow{1}{*}{Arp 299-B AT1}& \multirow{1}{*}{0.010411}& \multirow{1}{*}{Infrard}%{near-IR}
    & \multirow{1}{*}{2005 January 30}& \citet{Mattila2018-arp299}\\
\cmidrule(lr){1-5}
    \multirow{1}{*}{ASASSN-14ae}& \multirow{1}{*}{0.0436}& \multirow{1}{*}{Optical}& \multirow{1}{*}{2014 January 25}& \citet{Cendes2024}\\
\cmidrule(lr){1-5}
    \multirow{4}{*}{ASASSN-14li}& \multirow{4}{*}{0.0206}&\multirow{4}{*}{Optical}& \multirow{4}{*}{2014 November 22}& \citet{Alexander16}\\ & & & & \citet{Velzen2016}\\ & & & & \citet{Bright2018}  \\
    &&&&\citet{Anumarlapudi2024-radiosample}\\
\cmidrule(lr){1-5}
    \multirow{3}{*}{ASASSN-15oi}& \multirow{3}{*}{0.0484}& \multirow{3}{*}{Optical}& \multirow{3}{*}{2015 August 14}& \citet{Horesh2021-15oi}\\& & & &\citet{Hajela2025-15oi}\\
    &&&&\citet{Anumarlapudi2024-radiosample}\\
\cmidrule(lr){1-5}
    AT2018bsi&0.051&Optical&2018 April 9&\citet{Cendes2024}\\
\cmidrule(lr){1-5}
    \multirow{2}{*}{AT2018cqh}& \multirow{2}{*}{0.048}& \multirow{2}{*}{Optical}& \multirow{2}{*}{2018 June 16}&\citet{Zhang2024-18cqh}\\
    &&&&\citet{Yang2025ApJ-18cqh}\\
\cmidrule(lr){1-5}
    \multirow{1}{*}{AT2018dyb/ASASSN-18pg}& \multirow{1}{*}{0.018}& \multirow{1}{*}{Optical}& \multirow{1}{*}{2018 July 11}&\citet{Cendes2024} \\
\cmidrule(lr){1-5}
    \multirow{3}{*}{AT2018fyk/ASASSN-18ul}&\multirow{3}{*}{0.06} &\multirow{3}{*}{Optical} &\multirow{3}{*}{2018 September 8}&\citet{Wevers2019-AT2018fyk}\\
    &&&&\citet{Wevers2021-AT2018fyk}\\
    &&&&\citet{Cendes2024ATel-AT2018fyk}\\
\cmidrule(lr){1-5}
    \multirow{2}{*}{AT2018hco}& \multirow{2}{*}{0.088}& \multirow{2}{*}{Optical}& \multirow{2}{*}{2018 September 18}&\citet{Horesh2018-AT2018hco_radio} \\& & & &\citet{Cendes2024} \\
\cmidrule(lr){1-5}
    \multirow{6}{*}{AT2018hyz/ASASSN-18zj}& \multirow{6}{*}{0.04573}& \multirow{6}{*}{Optical}& \multirow{6}{*}{2018 October 14}&\citet{Gomez2020-AT2018hyz_radio_nodetect}\\& & & &\citet{Cendes2022-AT2018hyz}\\& & & &\citet{Sfaradi2024-AT2018hyz}\\& & & &\citet{Anumarlapudi2024-radiosample}\\
    &&&&\citet{Cendes2024}\\
    & & & &\citet{Cendes2026_AT2018hyz_radio}\\
\cmidrule(lr){1-5}
    \multirow{1}{*}{AT2018zr/PS18kh}& \multirow{1}{*}{0.071}& \multirow{1}{*}{Optical}& \multirow{1}{*}{2018 March 2}&\citet{Cendes2024} \\
\cmidrule(lr){1-5}
    \multirow{2}{*}{AT2019ahk/ASASSN-19bt}& \multirow{2}{*}{0.0262}& \multirow{2}{*}{Optical}& \multirow{2}{*}{2019 January 29}&\citet{Christy2024ASASSN-19bt} \\
    &&&&\citet{Anumarlapudi2024-radiosample}\\
\cmidrule(lr){1-5}
    \multirow{4}{*}{AT2019azh/ASASSN-19dj}& \multirow{4}{*}{0.022}& \multirow{4}{*}{Optical}& \multirow{4}{*}{2019 February 12}& \citet{Goodwin2022-19azh}\\& & & &\citet{Sfaradi2022-19azh}\\
    &&&&\citet{Anumarlapudi2024-radiosample}\\
    &&&&\citet{Burn2025-AT2019azh}\\
\cmidrule(lr){1-5}
    \multirow{5}{*}{AT2019dsg}& \multirow{5}{*}{0.0512}& \multirow{5}{*}{Optical}& \multirow{5}{*}{2019 April 9}& \citet{Stein2021}\\ & & & & \citet{Cannizzaro2021-19dsg} \\& & & &\citet{Cendes2021-19dsg}\\
    & & & &\citet{Mohan2022-19dsg}\\
    &&&&\citet{Cendes2024}\\
\cmidrule(lr){1-5}
    \multirow{1}{*}{AT2019ehz}& \multirow{1}{*}{0.074}& \multirow{1}{*}{Optical}& \multirow{1}{*}{2019 April 29}& \citet{Cendes2024}\\
\cmidrule(lr){1-5}
    \multirow{1}{*}{AT2019eve}& \multirow{1}{*}{0.081}& \multirow{1}{*}{Optical}& \multirow{1}{*}{2019 May 5}&\citet{Cendes2024}\\
\cmidrule(lr){1-5}
    \multirow{4}{*}{AT2019qiz}&\multirow{4}{*}{0.01513} &\multirow{4}{*}{Optical} &\multirow{4}{*}{2019 September 18} &\citet{Brien2019a-AT2019qiz}\\
    &&&&\citet{Brien2019b-AT2019qiz}\\
    &&&&\citet{Anumarlapudi2024-radiosample}\\
    &&&&\citet{Alexander2025-delayed_radio_sample}\\
\cmidrule(lr){1-5}
    \multirow{2}{*}{AT2019teq}& \multirow{2}{*}{0.0878}& \multirow{2}{*}{Optical}& \multirow{2}{*}{2019 October 20}& \citet{Cendes2024}\\
    &&&&this work\\
\cmidrule(lr){1-5}
    \multirow{2}{*}{AT2020afhd}& \multirow{2}{*}{0.027}& \multirow{2}{*}{Optical}& \multirow{2}{*}{2024 January 1\tnote{a}}& \citet{Wang2025-2020afhd}\\
    &&&&\citet{Christy2024TNS-2020afhd} \\
\cmidrule(lr){1-5}
    \multirow{2}{*}{AT2020mot}&\multirow{2}{*}{0.070}&\multirow{2}{*}{Optical}&\multirow{2}{*}{2020 June 14}&\citet{Liodakis2023-AT2020mot}\\
    &&&&\cite{Cendes2024}\\
\cmidrule(lr){1-5}
    \multirow{1}{*}{AT2020neh}& \multirow{1}{*}{0.062}& \multirow{1}{*}{Optical}& \multirow{1}{*}{2020 June 19}& \citet{Cendes2024}\\
\cmidrule(lr){1-5}
    AT2020nov&0.084&Optical&2020 June 27&\citet{Cendes2024}\\
\cmidrule(lr){1-5}
    \multirow{1}{*}{AT2020opy}& \multirow{1}{*}{0.159}& \multirow{1}{*}{Optical}& \multirow{1}{*}{2020 July 8}&\citet{Goodwin2023-AT2020opy} \\
\cmidrule(lr){1-5}
    AT2020pj&0.068&Optical&2020 January 2&\citet{Cendes2024}\\
\cmidrule(lr){1-5}
    \multirow{2}{*}{AT2020vdq}& \multirow{2}{*}{0.045}& \multirow{2}{*}{Optical}& \multirow{2}{*}{2020 October 4}&\citet{Somalwar2025-AT2020vdq} \\
    &&&&this work\\
\cmidrule(lr){1-5}
    \multirow{2}{*}{AT2020vwl}& \multirow{2}{*}{0.0325}& \multirow{2}{*}{Optical}& \multirow{2}{*}{2020 October 08}&\citet{Goodwin2023-AT2020vwl} \\ &&&& \citet{Goodwin2025-AT2020vwl}\\
\cmidrule(lr){1-5}
    AT2020wey&0.027&Optical&2020 October 8&\citet{Cendes2024}\\
\cmidrule(lr){1-5}
    \multirow{1}{*}{AT2022dbl}& \multirow{1}{*}{0.0284}& \multirow{1}{*}{Optical}& \multirow{1}{*}{2022 February 22}&\citet{Hinkle2024-AT2022dbl} \\
\cmidrule(lr){1-5}
    \multirow{2}{*}{AT2022dsb}& \multirow{2}{*}{0.0235}& \multirow{2}{*}{X-ray}& \multirow{2}{*}{2022 February 17}& \citet{Malyali2024-22dsb} \\
    &&&&\citet{Anumarlapudi2024-radiosample}\\
\cmidrule(lr){1-5}
    \multirow{1}{*}{AT2024tvd}& \multirow{1}{*}{0.04494}& \multirow{1}{*}{Optical}& \multirow{1}{*}{2024 August 25}& \citet{Sfaradi2025-AT2024tvd}  \\
\cmidrule(lr){1-5}
    \multirow{1}{*}{CNSS J0019+00}& \multirow{1}{*}{0.018}& \multirow{1}{*}{Radio}& \multirow{1}{*}{2015 March 21}& \citet{Anderson2020-CNSS0019}\\
% \cmidrule(lr){1-5}
    % \multirow{1}{*}{EP240222a}& \multirow{1}{*}{0.032}& \multirow{1}{*}{X-ray}& \multirow{1}{*}{22 February 2024}&\citet{JinEP2025}  \\
\cmidrule(lr){1-5}
    \multirow{1}{*}{AT2020wjw/eRASSt J234403-352640}& \multirow{1}{*}{0.1}& \multirow{1}{*}{X-ray}& \multirow{1}{*}{2020 November 28}&\citet{Goodwin2024-J2344} \\
\cmidrule(lr){1-5}
    \multirow{2}{*}{AT2019avd/eRASSt J082337+042302}& \multirow{2}{*}{0.029/0.028}& \multirow{2}{*}{Optical}& \multirow{2}{*}{2019 February 9}&\citet{Wang2023-19avd} \\
    &&&&\citet{Goodwin2025-22eROSAT_sample}\\
% \cmidrule(lr){1-5}
    % \multirow{1}{*}{eRASSt J043959.5-651403}& \multirow{1}{*}{0.15}& \multirow{1}{*}{}& \multirow{1}{*}{}&\citet{Goodwin2025-22eROSAT_sample}  \\
% \cmidrule(lr){1-5}
    % \multirow{1}{*}{eRASSt J063413-713908}& \multirow{1}{*}{0.13}& \multirow{1}{*}{}& \multirow{1}{*}{}&\citet{Goodwin2025-22eROSAT_sample}  \\
% \cmidrule(lr){1-5}
    % \multirow{1}{*}{eRASSt J110240-051813}& \multirow{1}{*}{$0.33^{+0.10}_{-0.09}$}& \multirow{1}{*}{}& \multirow{1}{*}{}&\citet{Goodwin2025-22eROSAT_sample}  \\
% \cmidrule(lr){1-5}
    % \multirow{1}{*}{eRASSt J143624-174105}& \multirow{1}{*}{$0.19^{+0.06}_{-0.03}$}& \multirow{1}{*}{X-ray}& \multirow{1}{*}{}&\citet{Goodwin2025-22eROSAT_sample}  \\
% \cmidrule(lr){1-5}
    % \multirow{1}{*}{eRASSt J164649-692539}& \multirow{1}{*}{0.017}& \multirow{1}{*}{}& \multirow{1}{*}{}&\citet{Goodwin2025-22eROSAT_sample}  \\
% \cmidrule(lr){1-5}
    % \multirow{1}{*}{eRASSt J190147-552200}& \multirow{1}{*}{0.058}& \multirow{1}{*}{}& \multirow{1}{*}{}&\citet{Goodwin2025-22eROSAT_sample}  \\
\cmidrule(lr){1-5}
    \multirow{1}{*}{eRASSt J011431-593654}& \multirow{1}{*}{0.16}& \multirow{1}{*}{X-ray}& \multirow{1}{*}{2020 November 24\tnote{b}}&\citet{Goodwin2025-22eROSAT_sample}  \\
\cmidrule(lr){1-5}
    \multirow{1}{*}{eRASSt J142140-295321}& \multirow{1}{*}{0.056}& \multirow{1}{*}{X-ray}& \multirow{1}{*}{2020 January 30\tnote{b}}&\citet{Goodwin2025-22eROSAT_sample}  \\
\cmidrule(lr){1-5}
    \multirow{1}{*}{eRASSt J143624-174105}& \multirow{1}{*}{$0.19^{+0.06}_{-0.03}$}& \multirow{1}{*}{X-ray}& \multirow{1}{*}{2022 February 18\tnote{b}}&\citet{Goodwin2025-22eROSAT_sample}  \\
\cmidrule(lr){1-5}
    \multirow{1}{*}{eRASSt J210858-562832}& \multirow{1}{*}{0.043}& \multirow{1}{*}{X-ray}& \multirow{1}{*}{2020 October 26\tnote{b}}&\citet{Goodwin2025-22eROSAT_sample}  \\
\cmidrule(lr){1-5}
    \multirow{1}{*}{FIRST J153350.8+272729}& \multirow{1}{*}{0.03243}& \multirow{1}{*}{Radio}& \multirow{1}{*}{1986 November 6–December 13}&\citet{Ravi2022-FIRSTJ1533} \\
\cmidrule(lr){1-5}
    \multirow{4}{*}{IGR J12580+0134}& \multirow{4}{*}{0.00411}& \multirow{4}{*}{X-ray}& \multirow{4}{*}{2011 January 6}& \citet{Irwin2015-IGR}\\ & & & & \citet{Yuan2016-IGR}\\ & & & & \citet{Perlman2017-IGR} \\
    &&&&\citet{Perlman2022-IGR}\\
\cmidrule(lr){1-5}
    \multirow{2}{*}{iPTF 16fnl}& \multirow{2}{*}{0.016328}& \multirow{2}{*}{Optical}& \multirow{2}{*}{2016 August 26}&\citet{Horesh2021-16fnl_radio} \\& & & &\citet{Cendes2024}\\
\cmidrule(lr){1-5}
    OGLE17aaj&0.116&Optical&2017 January 2&\citet{Cendes2024}\\
\cmidrule(lr){1-5}
    \multirow{1}{*}{PS16dtm / SN2016ezh}& \multirow{1}{*}{0.0804}& \multirow{1}{*}{Optical}& \multirow{1}{*}{2016 August 12}&\citet{Cendes2024} \\
\cmidrule(lr){1-5}
    SDSS J1115+0544&0.08995 &Infrared &2015 May 28\tnote{c} &\citet{Zhang2025-SDSSJ1115}\\
\cmidrule(lr){1-5}
    \multirow{1}{*}{Swift J0230+28}&\multirow{1}{*}{0.036}& \multirow{1}{*}{X-ray}&  \multirow{1}{*}{2022 June 22}& \citet{Guolo2024-Sw_J0230} \\
\cmidrule(lr){1-5}
    \multirow{1}{*}{WTP14adeqka}& \multirow{1}{*}{0.01895}& \multirow{1}{*}{Infrared}& \multirow{1}{*}{2015 June 20}&\citet{Golay2025-WTP14adeqka}  \\
\cmidrule(lr){1-5}
    \multirow{1}{*}{XMMSL1 J0740-85}& \multirow{1}{*}{0.0173}& \multirow{1}{*}{X-ray}& \multirow{1}{*}{2014 April 1}& \citet{Alexander2017-XMMSL1}\\
\cmidrule(lr){1-5}
    \multirow{1}{*}{eRASSt J045650.3-203750}& \multirow{1}{*}{0.077}& \multirow{1}{*}{X-ray}&  \multirow{1}{*}{2020 September 8}& \citet{Liu2024-eRJ0456}\\
\cmidrule(lr){1-5}
    AT2019aalc&0.0356&Optical&2019 January 22&\citet{Veres2024-AT2019aalc-multband_radio}\tnote{d}\\
% \cmidrule(lr){1-5}
%     \multirow{1}{*}{}& \multirow{1}{*}{}& \multirow{1}{*}{}&  \multirow{1}{*}{}& \\
\bottomrule
\insertTableNotes
\end{longtable}
\end{ThreePartTable}

\section{Spectral Fitting Results} \label{app:data}

\begin{figure*}[htp]
\center
\includegraphics[width=1.0\textwidth]{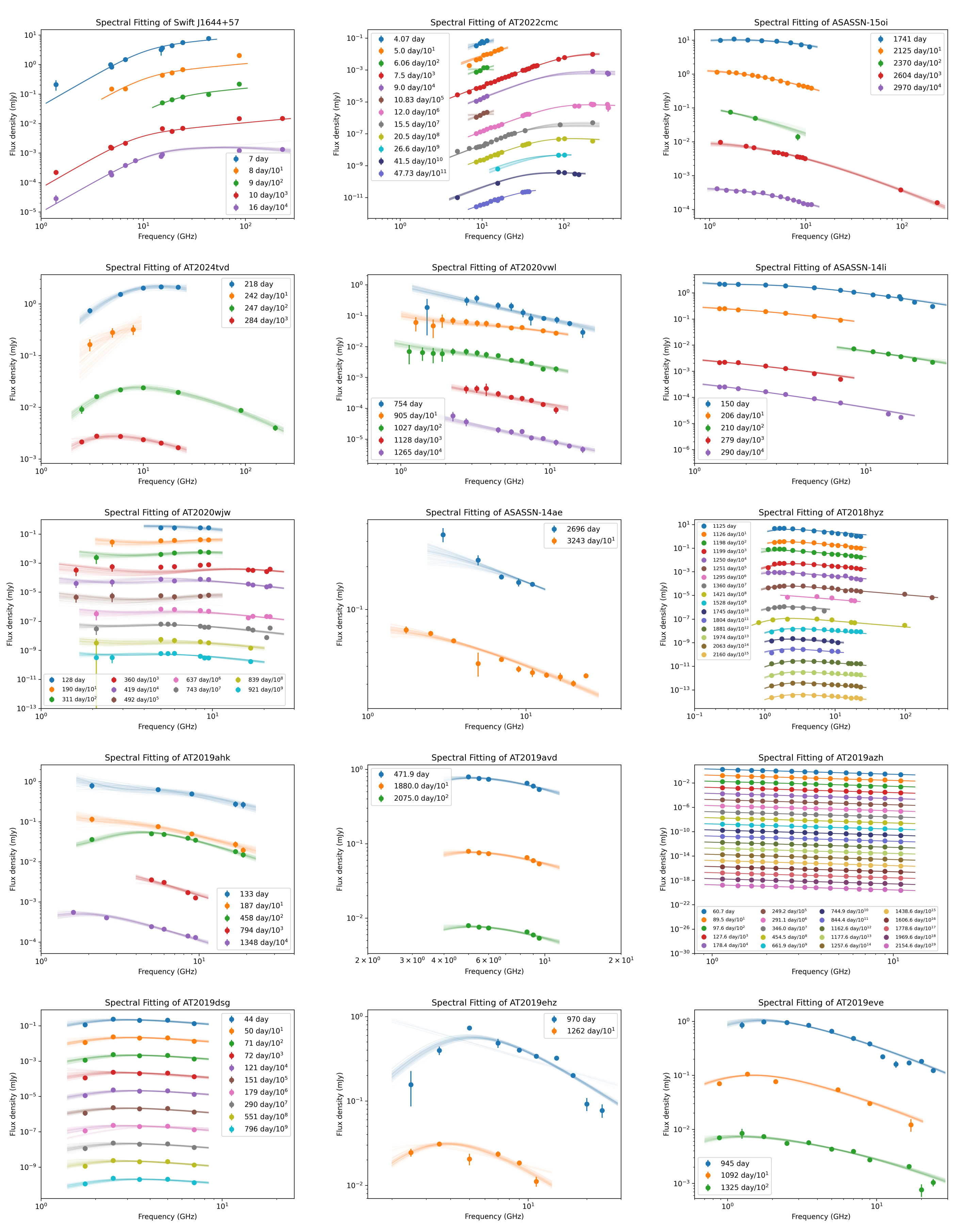}
\caption{Radio SED fitting results for 26 TDEs. The lines represent the fits from our MCMC modeling. }
\label{fig:sp_fit_1}
\end{figure*}

\begin{figure*}[htp]
\center
\includegraphics[width=1.0\textwidth]{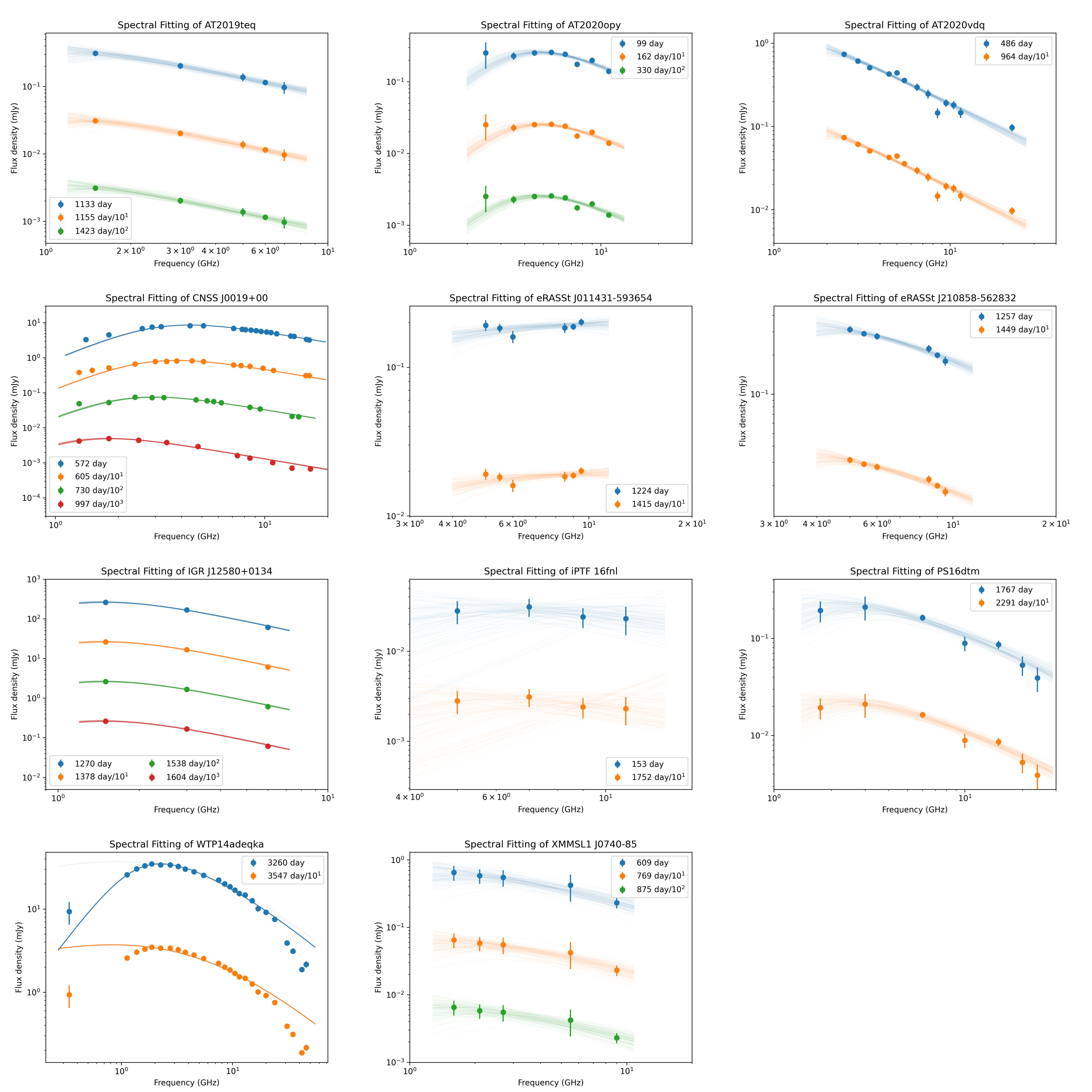}
\caption{Radio SED fitting results for 26 TDEs. The lines represent the fits from our MCMC modeling (continued). }
\label{fig:sp_fit_2}
\end{figure*}

\begin{figure*}[htp]
\center
\includegraphics[width=1.0\textwidth]{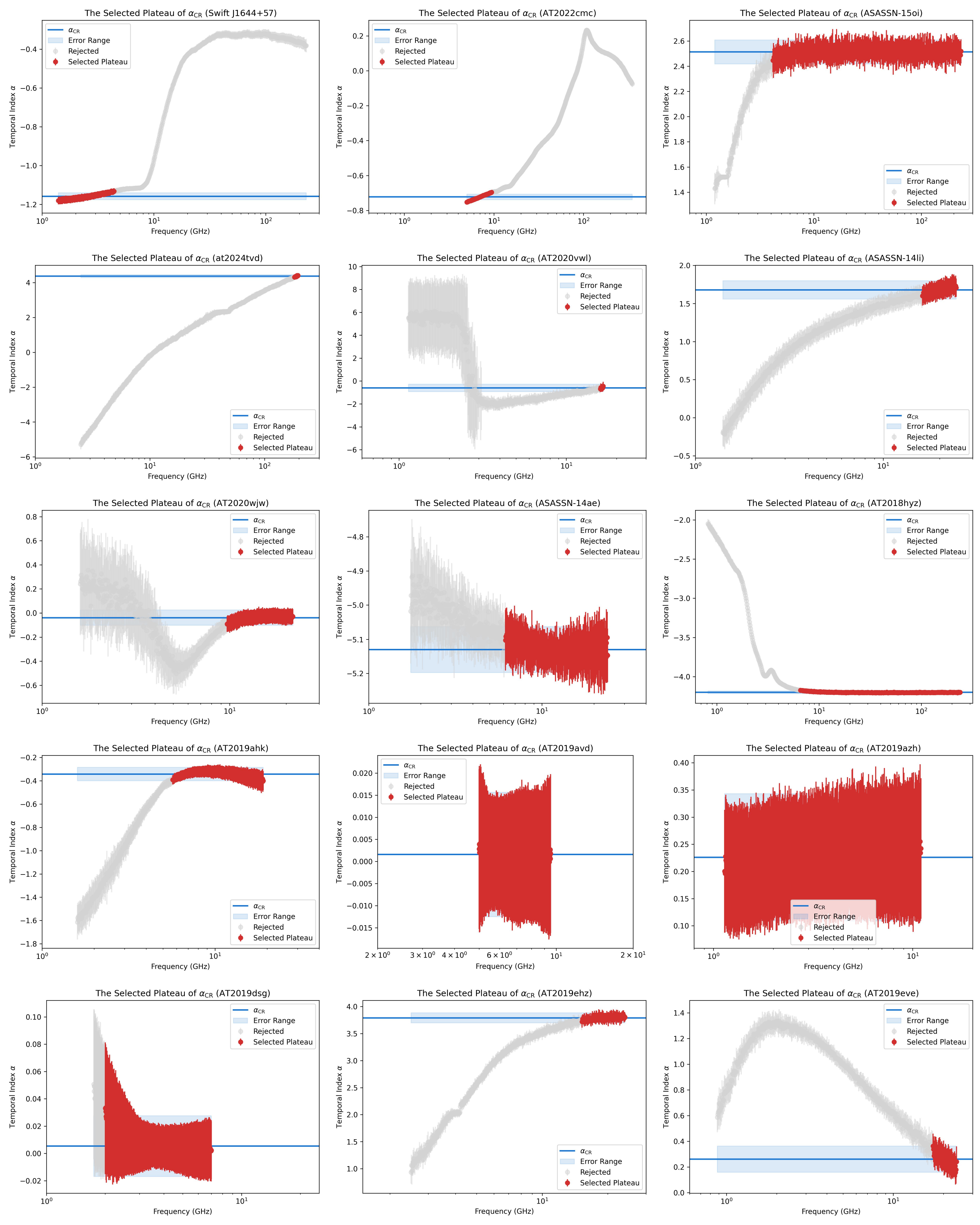}
\caption{The $\alpha(\nu) - \nu$ plane obtained with the power-law fit to the light curves of each frequency built from the spectral-fitting outputs. $\alpha(\nu)$ flattens lower frequency ($\nu< \min(\nu_{\rm a},\nu_{\rm m})$ and then $\beta_{\rm CR}=-2$) or higher frequency ($\nu> \max(\nu_{\rm a},\nu_{\rm m})$ and then $\beta_{\rm CR}=(p-1)/2$). Averaging $\alpha(\nu)$ over this plateau (red) gives $\alpha_{\rm CR}$ (the blue lines). The blue shaded regions show the error range.}
\label{fig:alpha_cr_1}
\end{figure*}

\begin{figure*}[htp]
\center
\includegraphics[width=1.0\textwidth]{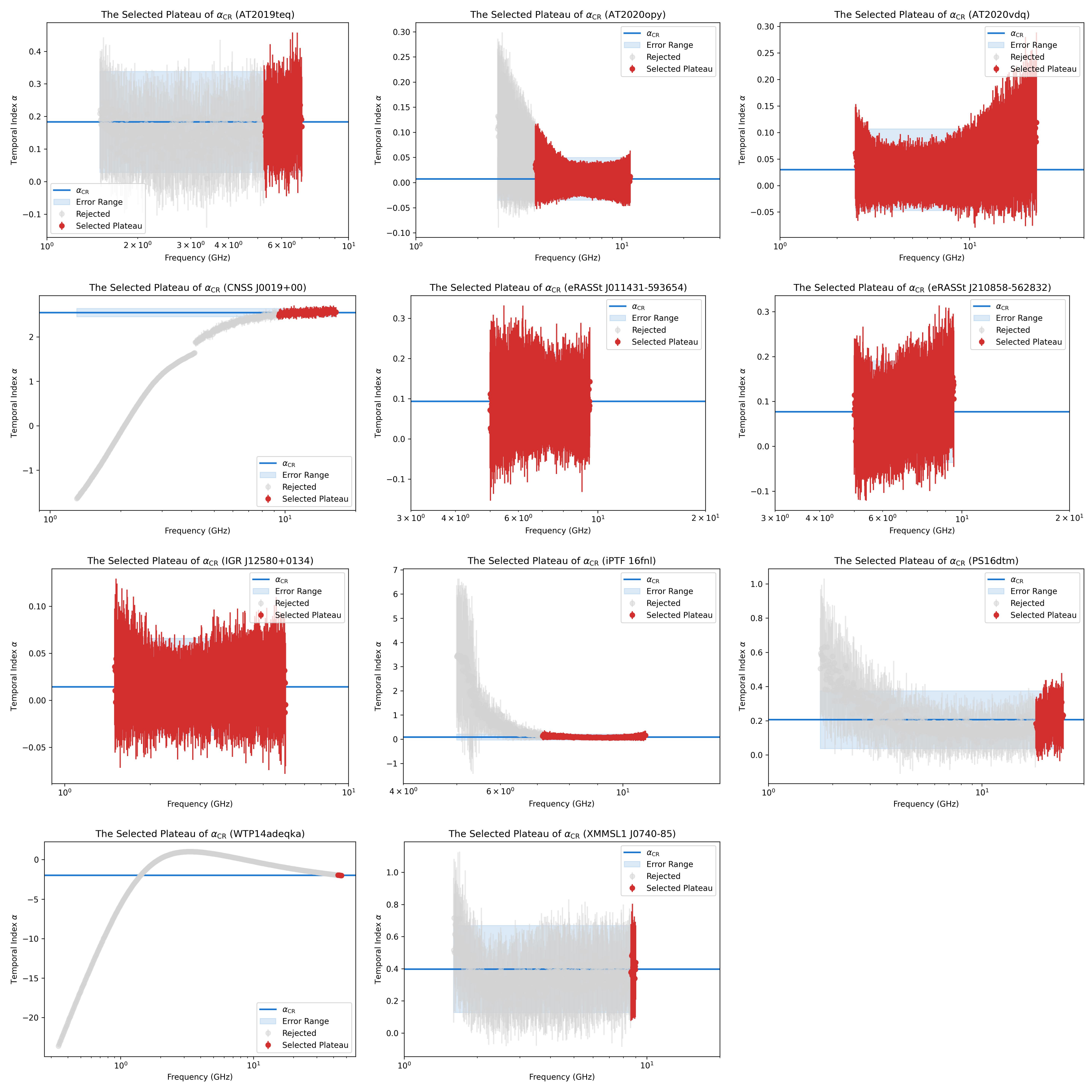}
\caption{The $\alpha(\nu) - \nu$ plane obtained with the power-law fit to the light curves of each frequency built from the spectral-fitting outputs. $\alpha(\nu)$ flattens lower frequency ($\nu< \min(\nu_{\rm a},\nu_{\rm m})$ and then $\beta_{\rm CR}=-2$) or higher frequency ($\nu> \max(\nu_{\rm a},\nu_{\rm m})$ and then $\beta_{\rm CR}=(p-1)/2$). Averaging $\alpha(\nu)$ over this plateau (red) gives $\alpha_{\rm CR}$ (the blue lines). The blue shaded regions show the error range (continued).}
\label{fig:alpha_cr_2}
\end{figure*}

\end{appendix}

\end{document}